\documentclass[letterpaper,twocolumn,10pt]{article}

\usepackage{usenix}
\usepackage[utf8]{inputenc}
\usepackage{graphicx}
\usepackage{amsmath}
\usepackage{xurl}
\usepackage{listings}
\usepackage{multirow}
\usepackage{algorithm}
\usepackage{algpseudocode}
\usepackage{pifont}
\usepackage{booktabs}
\usepackage{array}
\usepackage{subcaption}
\usepackage{paralist}
\usepackage{tabularx}
\usepackage[table]{xcolor}
\usepackage{authblk}
\usepackage[available]{usenixbadges}

\begin{document}

\date{}

\title{Security Analysis of LTE Connectivity in Connected Cars: A Case Study of Tesla}

\author{
Evangelos Bitsikas \quad Jason Veara \quad Aanjhan Ranganathan\\
Northeastern University, Boston, USA
}

\maketitle

\begin{abstract}
Modern connected vehicles rely on persistent LTE connectivity to enable remote diagnostics, over-the-air (OTA) updates, and safety-relevant services. While mobile network vulnerabilities are well documented in the smartphone ecosystem, their impact in safety-relevant automotive settings remains insufficiently examined. We conduct a black-box case study of LTE security in Tesla's Model 3 and Cybertruck, revealing systemic protocol weaknesses and architectural misconfigurations in connected vehicles. We find that Tesla's telematics stack is susceptible to IMSI catching, rogue base station hijacking, and insecure fallback mechanisms that may silently degrade service availability. Furthermore, legacy control-plane configurations allow for silent SMS injection and broadcast message spoofing without driver awareness. While the vulnerabilities are grounded in Tesla, this case study suggests broader implications for connected-vehicle telematics and for regulatory frameworks such as ISO/SAE 21434 and UN R155/R156, which assume secure, traceable, and resilient telematics in modern vehicles.

\end{abstract}

\section{Introduction}

The adoption of connected vehicles is accelerating rapidly, driven by the growth of intelligent transportation systems. These vehicles offer enhanced safety, convenience, and efficiency by maintaining continuous communication with cloud-based services~\cite{expertmarketresearch2025}. Models such as Tesla's Model 3 rely heavily on cellular networks to support remote diagnostics, over-the-air updates, infotainment, and safety-relevant functions like automatic crash notification (eCall)~\cite{TeslaModel3ServiceManual2024}[Sections 00, 21, ``Connectivity'' and ``In case of Emergency'']. As cellular integration deepens, the security and reliability of these communication channels become essential to ensure safe and uninterrupted vehicle operation.

Despite the adoption of mutual authentication, encryption, and integrity protection in LTE as specified by the 3rd Generation Partnership Project (3GPP), attacks such as IMSI catching, rogue base station exploitation, and signaling manipulation remain practical. These vulnerabilities have been demonstrated in numerous technical studies~\cite{Yang19:Overshadowing, Shaik15:PracticalAA, Kotuliak22:LTrack} and are periodically highlighted by real-world incidents reported in the media~\cite{Guardian2024, Wired2025, Australian2024}. Such attacks often bypass or exploit weaknesses in network implementation, policy enforcement, or device configuration, raising concerns for platforms that depend on persistent cellular access.

Securing connected vehicles introduces challenges that differ from those in conventional mobile platforms. Automotive telematics systems operate autonomously largely in a black-box, with minimal user interaction and limited visibility into network conditions. Unlike smartphones, where users can modify network settings, disable data services, or install monitoring tools, vehicles rely on OEM-defined network selection logic that is largely inaccessible to end users. This limits opportunities for in-the-loop detection and creates a broader attack surface. Furthermore, cellular connectivity in vehicles often supports critical features such as software updates, location reporting, and emergency services, increasing the potential impact of network-layer compromise. Several reports have raised concerns about Tesla's security posture~\cite{Reuters2025, InsideEVs2025, CarExpert2025}, alongside disclosures of non-cellular vulnerabilities~\cite{TeslaKeyCardVulnerability, TeslaRelayAttack, TeslaUWB, BluetoothAttackTeslaX}. Yet, the role of cellular misconfigurations and their concrete manifestations in connected vehicles remain underexplored.

This work presents the first security evaluation of cellular connectivity in connected vehicles, in the form of a Tesla-based case study centered on the Model 3 and Cybertruck. In other words, Tesla was selected primarily because it exposes more configurable interfaces and diagnostic endpoints than most commercial vehicles, enabling deeper access to network settings and behavior. This flexibility makes Tesla uniquely suitable for controlled experimentation and adversarial analysis, offering practical insight into real-world vehicular connectivity that is otherwise difficult to obtain. Our primary objectives are i) to assess whether connected vehicles exhibit LTE vulnerabilities known from other wireless devices, ii) characterize interoperability between the cellular modem and the vehicle software stack, iii) identify vehicle-specific behaviors that emerge from persistent connectivity, and iv) motivate closer scrutiny of cellular security practices in connected vehicles. Consequently, this case study bridges the gap between known cellular-layer threats and their practical realization in vehicular contexts by evaluating the system’s response to protocol-level attacks and adverse network conditions.

Specifically, we make the following contributions.

\begin{enumerate}

    \item \textbf{Automotive-Specific LTE Analysis Methodology.} We develop a custom LTE experimentation environment tailored to the constraints of connected vehicles, where network behavior is opaque, firmware is locked, and user interaction is minimal. By integrating a Tesla Model 3 and Cybertruck into our setup, we enable black-box, non-invasive testing that reveals how cellular threats manifest in vehicular contexts.

    \item \textbf{Empirical Analysis of Tesla's LTE Behavior.} We evaluate network-layer behaviors that influence resilience under degraded or adversarial conditions, including IMSI catching, false base stations, fallback loops, delayed recovery, and silent reception of SMS and emergency broadcasts. These subtle protocol-level failures can prolong vulnerability windows and degrade backend access without driver awareness.
    
    \item \textbf{Recommendations for Architectural Hardening.} We propose mitigations including stricter Public Land Mobile Network (PLMN) enforcement, rogue cell detection, and runtime safeguards against insecure network states. Though grounded in Tesla’s stack, these recommendations are also relevant to other OEMs and connectivity vendors that integrate similar cellular components and telematics architectures.

    \item \textbf{Regulatory Implications for R155 and R156.} We highlight how these vulnerabilities conflict with emerging automotive cybersecurity standards such as UN R155\cite{UNECE-R155} and R156~\cite{UNECE-R156}. Our findings show that LTE-layer attacks can silently undermine secure Over-the-Air (OTA) delivery, backend availability, and system transparency, raising compliance concerns in production vehicles.
\end{enumerate}

As part of our \textit{responsible disclosure} process, we shared all of our findings with Tesla~\footnote{\url{https://www.tesla.com/legal/security}}. In response, Tesla indicated that some of the observed issues may involve components in the cellular modem supplier stack, including Qualcomm and Quectel, in addition to the vehicle software layer. This suggests that the issues may not be unique to Tesla, but could affect a broader set of automotive OEMs that rely on similar cellular components and integration patterns. Consequently, our findings underscore systemic risks in connected vehicles, where risks involving shared cellular components or telematics integration patterns may propagate across manufacturers, thereby motivating regulatory and industry-wide mitigations. We subsequently followed up directly with Quectel~\footnote{\url{https://www.quectel.com/psirt-policy/}} and Qualcomm~\footnote{\url{https://www.qualcomm.com/company/product-security}}, sharing the complete set of findings and supporting results. Both vendors acknowledged receipt and indicated that they would investigate the reported issues further. To protect the vendors and minimize risks, we report only the appropriate supporting logs and evidence in the paper.

\section{Background and Related Work}

\textbf{Tesla's Connectivity Components.} Tesla’s connectivity framework integrates a dedicated \textit{Telematics Control Unit (TCU)}~\cite{TeslaModel3ServiceManual2024}[Section 2135], specialized firmware, and a set of cloud services providing reliable LTE-based communication for critical vehicle functions. The TCU itself is typically an embedded module that consolidates a cellular LTE modem, SIM/eSIM capabilities, and network management software. This module interfaces with both the vehicle’s internal networks (e.g., CAN bus) and external cellular networks, for data exchange between the vehicle and Tesla’s backend servers.  

By default, Tesla relies on preinstalled eSIMs rather than physical SIM cards, enabling carrier selection and roaming agreements without the need to physically insert a SIM card. Additionally, the \textit{On-Board Diagnostics (OBD)} system enables self-diagnosis and reporting, allowing access to various vehicle parameters. Finally, the \textit{Control Screen} serves as the primary interface for user interaction with the vehicle’s settings, with access to the \textit{``Service Mode''}, which reveals detailed information about the vehicle’s status.

\noindent\textit{Terminology:} Throughout this paper, we use the term \emph{TCU} when referring to Tesla’s telematics software and its integration logic with the \emph{modem} (i.e., baseband firmware).

\noindent\textbf{Existing Works on Tesla and LTE Security.} Surveys discussing general wireless attack methods, particularly in the context of vehicular networks, provide broad overviews of potential threats and mitigation strategies~\cite{dibaei2019overview, ELREWINI2020100214}. Notably, Foster \textit{et al.}~\cite{foster_woot15} studied an aftermarket TCU and showed that remote compromise of the device could enable arbitrary remote control of the vehicle, highlighting the broader security relevance of telematics.

Prior work has documented exploits on LTE protocols and control-plane messages, including capability manipulation and overshadowing attacks~\cite{jover16:ltesecurity, Shaik15:PracticalAA, Shaik19:Capabilities, Yang19:Overshadowing}. Adversaries can deploy rogue eNodeBs (Evolved Node B) to coerce attachments for interception, manipulation, or DoS~\cite{Shaik18:SoN, Rupprecht20:IMP4GT, Rupprecht19:Layer2, Bitsikas21:Handovers}, and force downgrades to weaker technologies (e.g., GSM)~\cite{Shaik19:Capabilities, karakoc23:Downgrades}. IMSI catchers exploit authentication gaps to elicit permanent identifiers, enabling tracking and profiling~\cite{Park19:AnatomyOC, Kotuliak22:LTrack, Yu19:Catchers, PALAMA21:IMSI-Catchers}. Persistent privacy risks also arise from paging, temporary identifiers, and capability exposure~\cite{Hong18:GUTI,oh24:Localization,Erni22:AdaptOver,Hussain19:Paging}. The SMS channel has also been abused for tracking, code execution, and DoS~\cite{Wen23:RILDefender, Mulliner09:SMS, Mulliner11:SMS, Bitsikas24:SMS, Bitsikas23:SMS}, and emergency alert systems have been shown vulnerable to manipulation~\cite{Lee19:Alerts, Bitsikas22:Warning}. 

Beyond individual attacks, several prior works have proposed adversarial and specification-driven cellular security testing frameworks that uncover protocol- and implementation-level flaws across devices, such as~\cite{Hussain18:LTEInspector, Kim19:Untouchables, bitsikas23:5g-framework, khandler24:astra}. Representative examples also include protocol-level and OTA-based testing/fuzzing systems, as well as frameworks that more directly target cellular protocol implementations or basebands~\cite{rupprecht16:ltefuzz, Park22:lte-test, park25:otabase, hoang25:llfuzz, garbelini22:autofuzz4g5g, maier20:basesafe}. These works are designed to expose protocol inconsistencies, robustness issues, or implementation flaws under adversarial or malformed-input settings. While such frameworks are designed to uncover both protocol- and implementation-level issues, our study focuses on protocol-compliant OTA testing and network-level behaviors observable in a black-box vehicular setting. In particular, we examine how a production vehicle TCU responds to adversarial but standards-compliant cellular conditions, such as rogue-cell selection, IMSI disclosure, fallback failures, PLMN handling, and silent handling of control-plane messages, rather than directly targeting the baseband or modem for software implementation flaws, such as memory-safety vulnerabilities, malformed-inputs, or remote code execution.

Moreover, these frameworks have been highly effective for smartphones and generic UEs, where researchers have sufficient control of the User Equipment (UE) (can readily swap SIM/eSIM profiles, control subscription parameters, and instrument modem behavior). However, transferring such approaches to \emph{connected vehicles} is non-trivial. In fact, vehicle TCUs are often provisioned with OEM-managed eSIM profiles and restrictive carrier policies, limiting researchers' ability to vary cellular settings or reproduce edge-case behaviors under controlled conditions. Observability is also reduced, with vehicles typically exposing only coarse connectivity status rather than cellular-layer diagnostics (e.g., reject causes or bearer/PDN failures), and safety/legal constraints limit experimentation (e.g., on-road testing). Finally, automotive-specific characteristics (e.g., autonomous recovery/fallback logic, backend service workflows, and user-invisible control/message channels) are not directly addressed by such frameworks, as they are largely outside their scope. Accordingly, the vulnerability space we cover is centered on protocol-compliant cellular attack surfaces and system behaviors at the telematics/network boundary, rather than low-level modem or firmware exploitation.

To our knowledge, \textit{our work} is the first security assessment on vehicle cellular connectivity through practical integration into a custom network, as a case study for Tesla vehicles. We therefore bridge the gap between cellular threats and real-world vehicular impact by testing Tesla’s telematics module against LTE threats.

\section{Experimental Methodology}

\noindent\textbf{Software \& Hardware.} We evaluate the Tesla in a fully controlled LTE setup using Amarisoft Callbox Classic~\cite{amarisoft_callbox} and srsRAN~\cite{srsRAN} with USRP B210 Software-Defined Radios (SDR)~\cite{ettus_b210}, giving complete control of the RAN-EPC for both standard-compliant signaling and targeted manipulations. For SIM provisioning, we use sysmocom sysmoISIM-SJA5 cards~\cite{sysmoisim_sja5} programmed via \texttt{sysmo-usim-tool} to vary PLMN entries, operator identifiers, and authentication profiles. Traffic and control traces are captured with \texttt{Wireshark}~\cite{Wireshark} and \texttt{tcpdump}~\cite{tcpdump}, and basic health checks use \texttt{ping}~\cite{ping}, \texttt{traceroute}~\cite{traceroute}, and \texttt{iperf}~\cite{iperf}. We also use a TP-Link AC1750 WiFi~\cite{tp-link} inside the tent as a fallback for further experimental observations.

\begin{figure}[!t]
    \centering
    \begin{subfigure}{0.48\columnwidth}
        \centering
        \includegraphics[width=\linewidth]{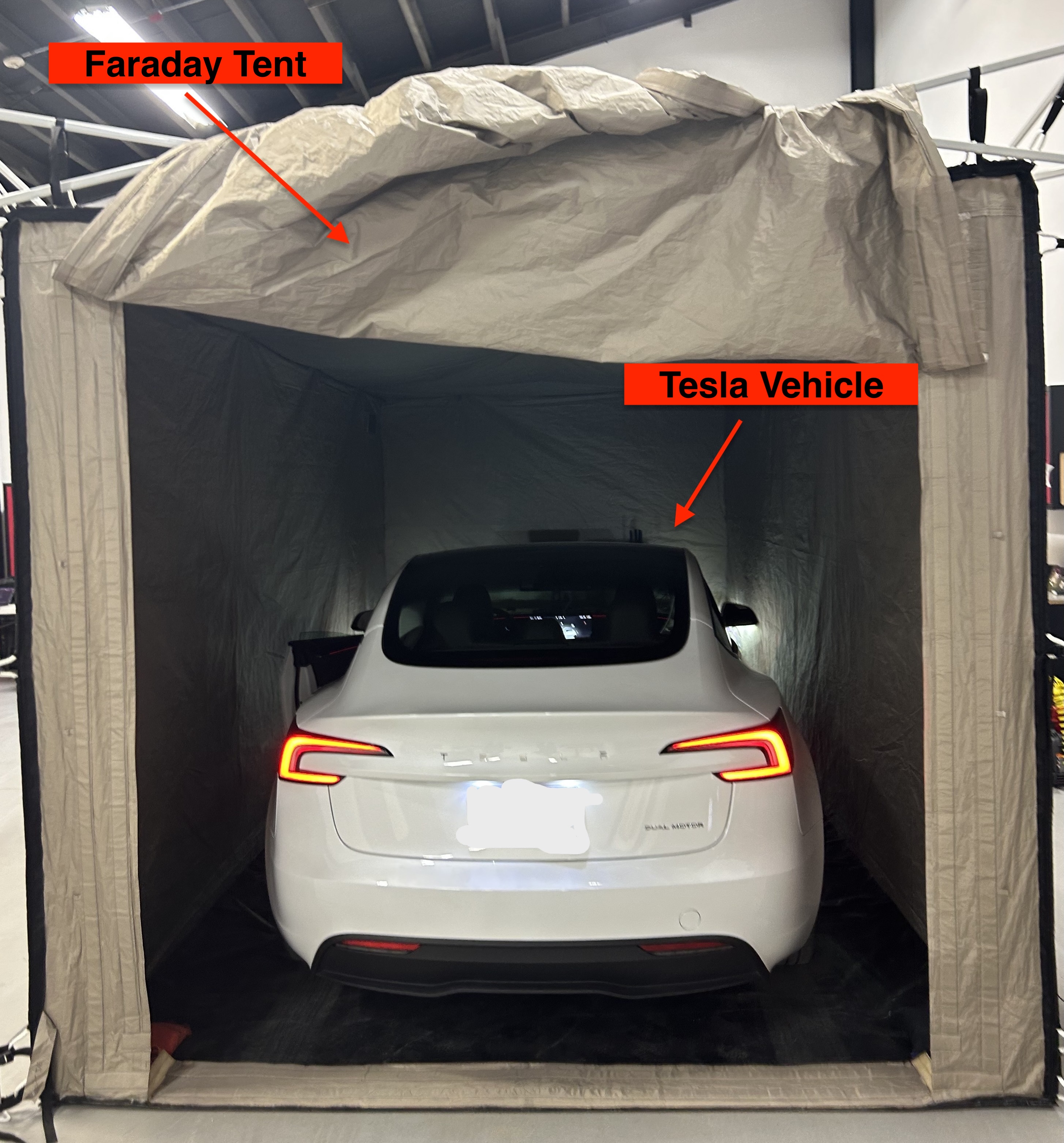}
        \caption{Vehicle \& Tent.}
        \label{fig:exp-car}
    \end{subfigure}
    \hfill
    \begin{subfigure}{0.48\columnwidth}
        \centering
        \includegraphics[width=\linewidth]{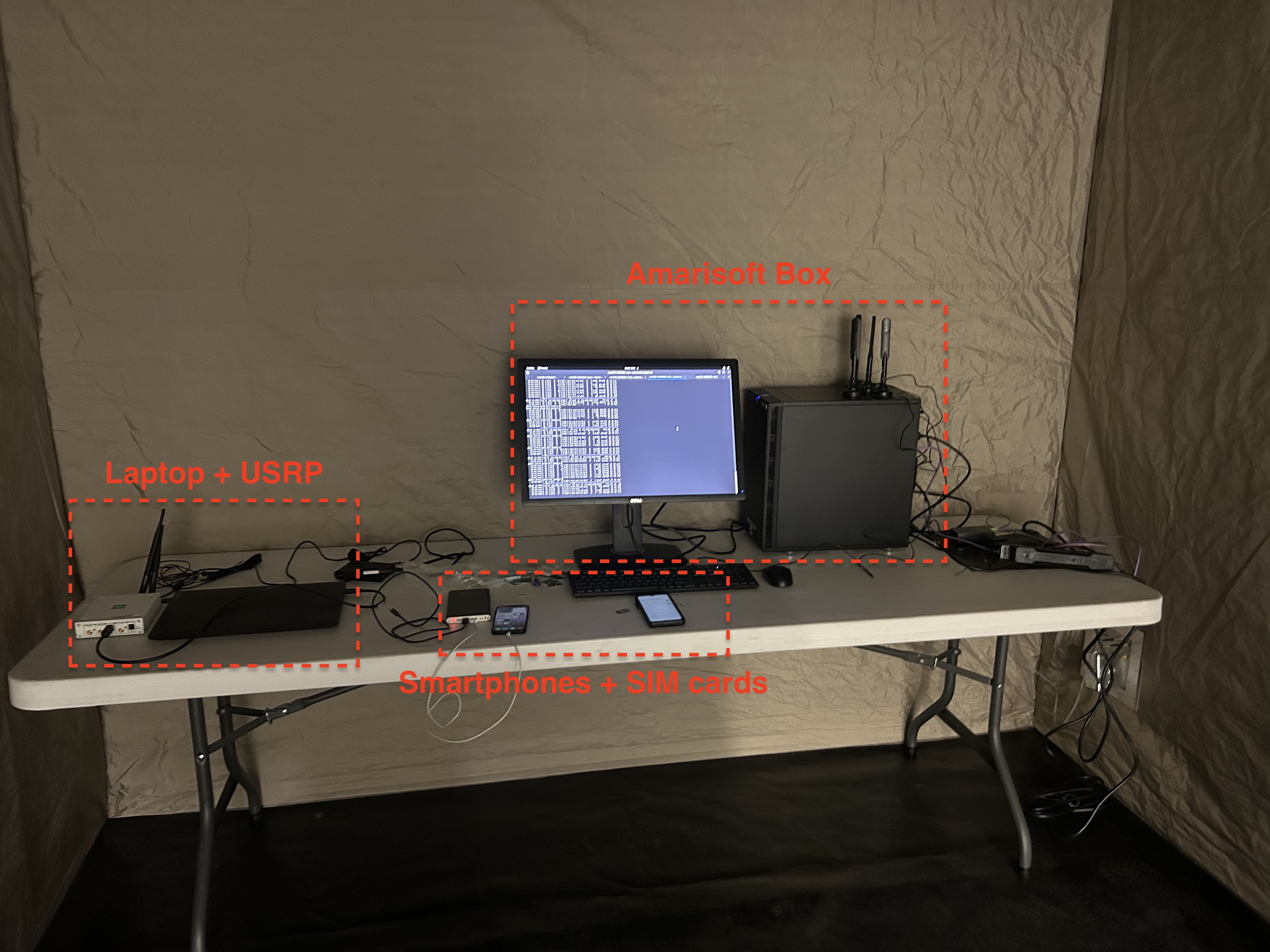}
        \caption{Equipment.}
        \label{fig:exp-equipment}
    \end{subfigure}
    \caption{The equipment for the LTE experiments.}
    \label{fig:exp-combined}
\end{figure}

\noindent\textbf{Shielded Testing.} All experiments ran inside a full-size Faraday tent~\cite{SelectFabricators} enclosing the vehicle, radios, and networking gear. The enclosure provides $\approx$ 93 dB RF isolation, preventing interference with commercial networks and blocking external signals, which ensures ethical, reproducible results. Figures~\ref{fig:exp-car} and~\ref{fig:exp-equipment} illustrate the experimental environment.

\noindent\textbf{Vehicles Under Test.} We evaluated a 2024 U.S. Tesla Model~3~\cite{TeslaModel3ServiceManual2024} and Cybertruck~\cite{TeslaCybertruckServiceManual2024}, each using a Quectel AG525R-GL modem~\cite{quectel_ag525r_gl} and a restricted Linux-based shell (diagnostics access). In Tesla service diagnostics, both vehicles reported the native LTE telematics unit as ``TELEMATICS CONTROL UNIT - TCU - LTE, 1702588-S0-I''. Limited OBD-over-Ethernet access supported authorized diagnostics and telemetry via the Tesla Toolbox~\cite{TeslaToolbox}. We performed strictly \textit{non-invasive tests} (no rooting, software bypasses, or credential extraction). All connectivity was orchestrated through our own LTE setup to model an external wireless adversary under realistic constraints. We utilized the physical SIM slot for further security analysis only, as it is normally not a requirement for wireless attacks.

\noindent\textbf{Establishing Controlled LTE Connectivity.} The TCU defaults to a commercial eSIM profile (e.g., T-Mobile/AT\&T) and prioritizes it over the physical SIM unless explicitly reconfigured via Tesla Toolbox. To operate in a controlled network, we combined passive observation, programmable SIMs, and custom RAN-EPC configurations. Baseline monitoring showed strong eSIM preference, continuous PLMN scanning, and concurrent cellular–WiFi use. We then mapped the local RF environment (CellMapper, srsRAN, Android dialer~\cite{CellMapper,srsRAN,AndroidDialer}) to strategically select carriers, bands and parameters. We provisioned programmable SIMs to replicate target PLMNs and set the authentication/APN parameters (e.g., Ki, OPC, APN) accordingly. The LTE setup was validated with reference smartphones, before the vehicle was introduced and the TCU connected successfully. Through Tesla Toolbox and Service Mode, we verified SIM recognition, network registration, and active data sessions, confirming that our controlled LTE environment could support reliable connectivity for subsequent security evaluations. These capabilities allowed us to recreate a wide array of realistic adversarial LTE behaviors without breaching ethical constraints. The LTE architecture is further covered by Appendix Section~\ref{sec:architecture}, whereas Sections~\ref{sec:equipment} and~\ref{sec:method} provide further information about the vehicle, equipment and the connection methodology.

\textbf{Ethics.} Given our careful experimental design, all tests were conducted in a controlled and non-invasive environment, thereby minimizing unnecessary risks. As noted in the Introduction, we also followed a responsible disclosure process and shared all findings with the affected parties.

\section{Cellular Security Risks}
\label{sec:risks}

Below, we describe six cellular security aspects identified and analyzed in the context of connected vehicles. The scenarios described here were selected based on their relevance to real-world automobiles, feasibility under our testbed conditions, and impact on security, connectivity, or network trust. While not exhaustive, they represent a set of practically verifiable risks that serve as the basis for the empirical analysis.

\textbf{Threat Model.} We consider a purely wireless adversary and distinguish two attacker classes. \textit{(A1) Passive wireless observer:} an adversary that can monitor LTE broadcasts, configurations, unencrypted data and exposed identifiers, but does not transmit or alter network behavior. \textit{(A2) Active rogue base station attacker:} an adversary with wireless access only, who can operate rogue LTE eNodeBs (or transceivers) in proximity to the vehicle, spoof broadcast identifiers (e.g., PLMN and cell parameters), and manipulate standards-compliant pre-authentication and control-plane signaling to influence cell selection, trigger identity disclosure, induce incomplete attachment, or degrade connectivity. This attacker has no access to internal vehicle software, OEM backend infrastructure, physical hardware, or long-term SIM credentials.

The experiments in this paper are grounded in these wireless attacker models. For some evaluations, we use a controlled LTE setup and programmable SIM/network parameters to safely reproduce, observe, and diagnose system behaviors that can assist or amplify such wireless attacks. These additional controls are used as experimental instrumentation and do not imply a separate insider or privileged attacker model unless explicitly stated. Table~\ref{tab:threat_mapping} maps each finding to the wireless attacker class most directly associated with it and clarifies whether it constitutes a direct attack, an enabling condition, or a system property that assists or amplifies such an attacker.

\subsection{Classic Cellular Attacks}

Connected vehicles inherit many of the same cellular-layer vulnerabilities in mobile networks, but their impact in automotive settings is less well understood. In contrast to mobile phones, vehicles operate with continuous connectivity, minimal user interaction, and rigid network logic. This combination can make certain attacks more feasible or damaging. This section focuses on two foundational LTE threats, IMSI catching and rogue base station exploitation. These were selected not for historical relevance alone, but because they remain practically persistent under realistic adversarial models. 

\noindent\textbf{$\bullet$ IMSI Catching.} The International Mobile Subscriber Identity (IMSI) is a globally unique identifier assigned to every mobile subscriber in 3GPP-based networks such as LTE~\cite{3gpp.23.003}. To reduce the risk of tracking, devices typically use temporary identifiers, such as the Temporary Mobile Subscriber Identity (TMSI). However, during certain control-plane procedures, particularly initial registration, network reattachment, or when temporary identifiers are unavailable, a device may need to transmit its IMSI in cleartext.

Beyond passive eavesdropping, active attackers run rogue eNodeBs that spoof PLMNs (often at higher power) to trigger attachment, then coerce identifier disclosure (e.g., pre-authentication Non-Access Stratum (NAS) \texttt{IdentityRequest}). Because 3GPP permits networks to request the real IMSI for administrative/authentication purposes before mutual authentication, UEs may comply, enabling tracking, denial-of-service, and follow-on protocol attacks. Such cases on phones are well documented~\cite{PALAMA21:IMSI-Catchers, Yu19:Catchers, Erni22:AdaptOver, Kotuliak22:LTrack, Park19:AnatomyOC}.

However, connected vehicles present elevated concerns. Automotive TCUs can rely on LTE connectivity to support remote diagnostics, OTA updates, and application communications. Unlike smartphones, which may connect intermittently and allow user intervention, TCUs automatically reattach to available LTE cells, often prioritizing signal strength over authenticity. This behavior increases their exposure in environments such as parking structures, garages, or dense urban areas where attackers operate multiple rogue eNodeBs. Telematics units use pre-provisioned eSIMs, lack user-accessible network settings, and their operating systems are typically locked down and managed remotely. In contrast, smartphones allow users to change carriers, disable data, activate airplane mode, or use mobile applications to detect rogue activity, like in~\cite{Tyler25:IMSI, Dabrowski14:IMSIcatch} (even as rooted/jail-broken). Even with a physical SIM active, fallback behavior is opaque and controlled entirely by the OEM, without on-screen alerts. The absence of user-facing diagnostics, observation and control makes this a persistent and underaddressed risk in automotive systems.

\begin{figure}[!t]
     \centering
     \includegraphics[width=0.9\columnwidth]{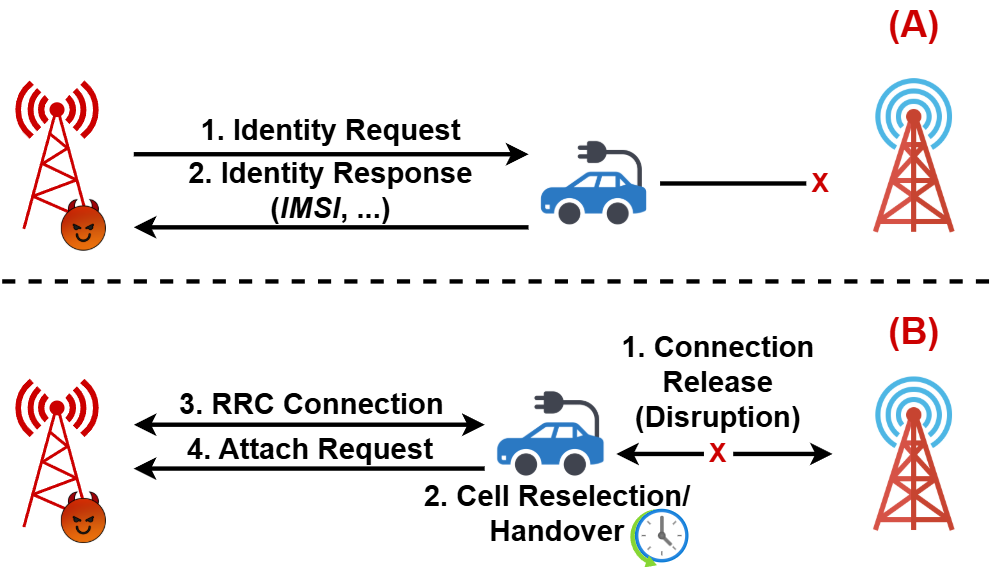}
     \caption{LTE attack examples: (A) IMSI Catching and (B) False Base Stations.}
     \label{fig:lte-attacks}
\end{figure}

\noindent\textbf{$\bullet$ False Base Stations.} Unlike IMSI catchers, which elicit identifiers without full registration, FBSs emulate legitimate infrastructure to induce incomplete or full attachment, enabling control-plane manipulation, denial-of-service, and potential interception. The attacker configures parameters, such as EARFCN, PLMN ID, and cell identity, to mimic a legitimate carrier. By transmitting valid Master and System Information Blocks (MIBs and SIBs) over a strong downlink signal, the rogue eNodeB can provoke the target UE into connecting. Subsequent Radio Resource Control (RRC)/Non-access stratum (NAS) exchanges provide visibility and, in many cases, injection opportunities. Although LTE specifies mutual authentication, re-attach/fallback paths and implementation flaws can still permit unauthorized connections or negotiate weak settings (e.g., integrity-free modes). By sustaining the session, an FBS can prolong attachment, create MitM conditions, or disrupt procedures, as examined across cellular generations~\cite{jover16:ltesecurity,Shaik18:SoN,Rupprecht19:Layer2,Bitsikas21:Handovers} on smartphone devices.

Because TCUs maintain always-on LTE and expose minimal user control, a stronger rogue cell can force selection by detaching the car from its operator. Once attached, the attacker can manipulate control-plane signaling to degrade or block services, sustain unauthorized sessions, and interfere with remote access, diagnostics, or OTA, typically without driver visibility or alerts. While smartphones are also susceptible, the stakes are higher in vehicles. TCUs run autonomously, integrate with safety-relevant functions (e.g., eCall, telemetry, remote access), and receive updates. Thus, FBS attacks threaten not only data confidentiality but also availability and control, risking operational failures.

\subsection{Security-Relevant System Behaviors}

While cellular-layer attacks target specific protocol weaknesses, the way a vehicle reacts to unexpected or degraded network conditions also affects its overall resilience. These system behaviors influence whether attacks succeed, how long their effects persist, and whether the vehicle can recover connectivity. We focus on four classes of behaviors that have both security and operational consequences. These areas were selected based on their known importance in LTE networks and their heightened impact in connected vehicles. These behaviors are not framed as requiring a different attacker model, but rather they characterize system properties that can increase the feasibility, persistence, or impact of the wireless attackers defined above.

\noindent\textbf{$\bullet$ Fallback Behavior and Partial Attach States.} Fallback behavior determines how a vehicle responds to rejected operators, failed authentication, and routing issues. Telematics units that do not properly transition to alternative access technologies, such as 3G, or WiFi, may remain in a degraded or disconnected state. In many cases, devices may enter a persistent reattachment loop, repeatedly attempting to join a network without success, thereby increasing the attack surface and service downtime.

Further, we define a partial/camped state as a condition in which a UE completes only a subset of the connectivity stages required for service, e.g., it may achieve control-plane registration/attach signaling, but fails to fully establish the data session, i.e. Packet Data Network (PDN) connectivity, and end-to-end IP connectivity. These conditions may arise due to core-side failures, such as PDN rejection, incomplete bearer setup or routing failures. In such cases, the TCU must correctly detect and recover from the degraded state. Failure to do so can lead to prolonged unavailability or increased vulnerability to rogue base stations and downgrade attacks. By contrast, an ``attached'' state refers to successful registration and association with the network.

\noindent\textbf{$\bullet$ Operator Whitelisting and PLMN Handling.} Mobile devices often rely on a predefined list of authorized networks to prevent attachment to rogue base stations. For connected vehicles, the enforcement of such Public Land Mobile Network (PLMN) whitelisting is especially important given the lack of user interaction or real-time alerting. Weak or inconsistent enforcement could allow an adversary to configure a malicious setup that imitates a legitimate operator, causing the TCU to attach to an unauthorized network.

In this context, whitelisting refers not only to filtering of PLMN identifiers, but also to how the telematics unit interprets network selection rules during attachment and roaming. Improper handling may permit access to networks outside of the permitted roaming zones, or fail to validate changes in operator identity across layers such as PLMN, Tracking Area Code (TAC), and Evolved Absolute Radio Frequency Channel Number (EARFCN). These weaknesses could be used to initiate unauthorized connectivity, inject malformed configurations, or mislead the vehicle into accepting/processing untrusted control messages.

\noindent\textbf{$\bullet$ Control Plane and Capability Parameters.} The negotiation of encryption and integrity algorithms during network attachment is a fundamental part of 3GPP security. Devices are expected to reject insecure or unsupported modes, such as the null encryption option \texttt{EEA0}, to prevent passive or active interception. However, implementation choices and operator configurations may create gaps between standard compliance and actual behavior. Weak enforcement of ciphering policies during the NAS security setup, or the acceptance of fallback to insecure modes, can expose the vehicle to confidentiality and integrity violations. 

Additionally, support of legacy services, acceptance of invalid NAS-RRC elements and inadequate security checks can further degrade the vehicle's security posture. Differences in security handling across PLMNs, particularly under roaming or degraded network conditions, may complicate this risk too. These behaviors can also enable downgrade attacks or serve as preconditions for man-in-the-middle manipulation. Unlike smartphones, where users can actively monitor network anomalies (e.g., disable 2G) and promptly install critical updates, automotive TCUs typically operate autonomously, provide minimal visibility into network behaviors, and have slower, manufacturer-controlled update cycles, thereby prolonging exposure to known flaws.

\noindent\textbf{$\bullet$ Injection of SMS and Emergency Messages.} SMS remains a supported feature in modern LTE networks and can be delivered via IMS or through circuit-switched fallback (CSFB) paths. Certain deployments permit non-interactive variants such as flash, silent, or binary SMS, which may carry hidden payloads intended to reconfigure device parameters, request data, or trigger abnormal behavior. These message types can be exploited by adversaries to inject deceptive service notifications or control signals, especially if the recipient device lacks adequate validation or filtering mechanisms. Prior works have demonstrated such risks across platforms, including inter-UE attacks using intercarrier routing paths with limited safeguards~\cite{Wen23:RILDefender, Mulliner09:SMS, Mulliner11:SMS, Bitsikas23:SMS, Bitsikas24:SMS, foster_woot15}.

Broadcast emergency alerts in LTE are similarly vulnerable. These alerts are transmitted via the Cell Broadcast System (CBS) using unprotected System Information Blocks (SIBs), such as SIB10 for earthquake warnings, SIB11 for extreme weather, and SIB12 for other alerts. Delivery is typically triggered through paging messages, prompting user equipment to retrieve the associated SIB content. An attacker with a rogue base station can spoof such messages by transmitting at higher signal strength, deceiving nearby devices into displaying falsified alerts. Alternatively, a broadcast alert can be suppressed by jamming or disrupting the paging cycle. Attacks of this nature have been shown to cause misinformation or failure to deliver time-sensitive warnings~\cite{Bitsikas22:Warning, Lee19:Alerts, Hussain19:Paging, Yang19:Overshadowing}.

In vehicles, the risks of SMS and emergency broadcast manipulation are amplified. Telematics systems often operate without user interaction and may process incoming messages silently. Unlike smartphones, which present alerts to the user and allow for filtering or configuration, TCUs may act on SMS or CBS data without visibility or manual override. This lack of transparency makes it difficult to detect or mitigate unauthorized messages. An injected SMS or spoofed emergency alert could trigger alarms, disrupt services, or activate safety-related failover logic. Since drivers are not notified and cannot disable such behavior, the potential impact can extend to service degradation and safety-critical malfunction.

\section{Empirical Evaluation and Results} \label{sec:results}

\begin{figure}[!t]
    \centering
    \begin{subfigure}{\columnwidth}
        \centering
        \includegraphics[width=\linewidth]{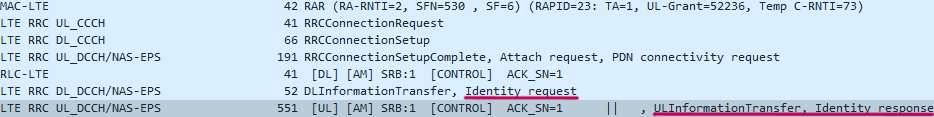}
        \caption{Identity request and response exchange.}
        \label{fig:identity-req-res}
    \end{subfigure}

    \vspace{0.0em}

    \begin{subfigure}{\columnwidth}
        \centering
        \includegraphics[width=\linewidth]{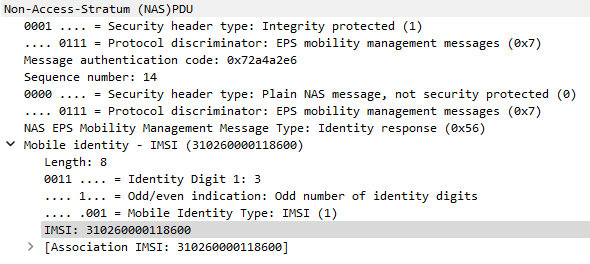}
        \caption{IMSI exposure via unprotected NAS signaling (pSIM).}
        \label{fig:imsi-exposure}
    \end{subfigure}

    \caption{Example of IMSI catching against Tesla Model 3.}
    \label{fig:imsi-catching}
\end{figure}

\noindent\textbf{$\bullet$ IMSI Catching Exposure.} To evaluate the susceptibility to IMSI catching, we conducted a series of rigorous experiments using both its eSIM (pre-provisioned with a T-Mobile profile) and a custom-programmed physical SIM (imitating a T-Mobile network). Our attacks were based on active LTE identity capture techniques involving the deployment of a rogue eNodeB that sends unprotected NAS \texttt{IdentityRequest} messages with spoofing, and with incomplete attachments to elicit the device's IMSI. Figure~\ref{fig:imsi-catching} shows an experimental example of an \texttt{IdentityRequest} message during such an attachment, with sub-Figure~\ref{fig:identity-req-res} showing the identity request, while Figure~\ref{fig:imsi-exposure} illustrates the exposed IMSI.

We conducted controlled IMSI catching experiments using both Amarisoft and srsRAN. The stations broadcast high-priority PLMNs, either legitimate identifiers (e.g., T-Mobile's \texttt{310260}) or non-commercial test codes (e.g., \texttt{00101}, \texttt{99970}), on the appropriate frequency band (primarily LTE Band~2). We tested both the eSIM and a custom-programmed physical SIM (pSIM) across 10 independent trials per configuration. Our results indicate a high degree of IMSI catching susceptibility under specific conditions. Table~\ref{tab:imsi-catching-summary} summarizes the outcomes of our IMSI catching experiments across various attacker PLMN configurations and SIM types. The Tesla TCU disclosed both permanent (IMSI) and globally unique temporary (GUTI) identifiers when exposed to rogue eNodeBs with matching PLMN values. Conversely, when test PLMNs (e.g., \texttt{00101}, \texttt{99970}) were used, the TCU refused to engage with the rogue network, indicating some basic filtering at the network selection stage. These findings confirm that IMSI disclosure is tightly linked to the use of legitimate PLMN values, even in adversarial conditions. It should be noted that the temporary identifiers may also introduce a potential privacy risk even without direct IMSI leakage, if not randomized and assigned frequently, leading to tracking if persistently correlated over time. However, we found no instances in which the IMEI was exposed via the NAS \texttt{IdentityRequest} message. While some implementations on mobile devices erroneously respond with the IMEI, this model appears to conform to proper identifier separation as per 3GPP.

\begin{table}[!t]
\centering
\small
\caption{IMSI catching outcomes with different PLMNs.}
\label{tab:imsi-catching-summary}
\renewcommand{\arraystretch}{1}
\rowcolors{2}{gray!10}{white}
\begin{tabular}{>{\centering\arraybackslash}p{1.4cm} 
                >{\centering\arraybackslash}p{1.2cm} 
                >{\centering\arraybackslash}p{1.7cm} 
                >{\centering\arraybackslash}p{1.1cm} 
                >{\centering\arraybackslash}p{1cm}}
\toprule
\rowcolor{gray!20}
\textbf{Attacking PLMN} & \textbf{Tesla PLMN} & \textbf{Identity Leaked} & \textbf{SIM Type} & \textbf{Attack Result} \\
\midrule
\midrule
310260 & 310260 & IMSI/GUTI & eSIM & \textcolor{green!60!black}{\ding{51}} \\
310260 & 310260 & IMSI/GUTI & pSIM & \textcolor{green!60!black}{\ding{51}} \\
310150 & 310150 & IMSI/GUTI & pSIM & \textcolor{green!60!black}{\ding{51}} \\
\midrule
00101   & 310260 & None        & pSIM & \textcolor{red}{\ding{55}} \\
99970   & 310260 & None        & pSIM & \textcolor{red}{\ding{55}} \\
\bottomrule
\end{tabular}
\end{table}

Overall, our findings demonstrate that Tesla’s cellular stack lacks effective defenses against IMSI catching attacks. Despite architectural and usage differences from smartphones, the vehicle's TCU remains equally susceptible to identity disclosure when exposed to adversarial signal conditions. This reinforces the broader systemic flaw in the cellular ecosystem, where transmitters are not authenticated and receivers do not prioritize trust, enabling adversaries to coerce devices into leaking sensitive identifiers without detection.

\noindent\textbf{$\bullet$ False Base Station Susceptibility.} We performed a series of experiments using our configured eNodeBs designed to operate a legitimate LTE infrastructure. Our primary objective was to assess whether Tesla’s TCU can be coerced into attaching to unauthorized LTE networks under realistic adversarial conditions, in which the vehicle was operational. We also deployed one or more rogue cells in a controlled RF environment, spoofing PLMN information with elevated signal power to outcompete the legitimate cells. 

During the testing, we varied parameters such as signal strength, broadcast band, tracking area code (TAC), and timing advance to examine how the TCU responded to subtle inconsistencies. Metrics collected included attach success rates, number of retries before fallback, and duration of control-plane engagement. Our logs showed that the TCU frequently entered an incomplete attach state when connecting to the rogue cell with spoofed carrier information. By matching the PLMN of Tesla’s native physical SIM, we were able to override the cell selection process and induce the TCU to prioritize our rogue eNodeB. Test PLMNs did not succeed in establishing a connection, which further confirms that the attacker needs to imitate the real network. By closely examining the control-plane logs (e.g., for \texttt{RRCConnectionReconfiguration}) and the subsequent attach attempts, we determined that the TCU not only accepted these configurations but also continued to exchange control-plane messages with rogue nodes, indicating also that it did not revert to a known trusted cell. Notably, the driver receives no visual or audible notification of the transition, making these attacks more stealthy.

\begin{table}[!t]
\centering
\caption{FBS result metrics across various cell gain values.}
\label{tab:fbs_vs_gain}
\renewcommand{\arraystretch}{1.1}
\setlength{\tabcolsep}{5pt} 
\begin{tabular}{lcccc}
\toprule
\multirow{2}{*}{\textbf{Metric}} & \multicolumn{4}{c}{\textbf{DL cell gain (dB)}}\\
& \textbf{0} & \textbf{-5} & \textbf{-10} & \textbf{-15}\\
\midrule
\rowcolor{gray!10}\textbf{Attach type} & Partial & Partial & Partial & Partial \\
\textbf{Time-to-attach (s)} & $\approx$ 2.0 & $\approx$ 2.0 & $\approx$ 2.5 & $\approx$ 4.5 \\
\rowcolor{gray!10}\textbf{Conn. duration (s)} & $\approx$ 4.0 & $\approx$ 4.0 & $\approx$ 4.0 & $\approx$ 4.0 \\
\textbf{Reattach / Camp} & \textcolor{red}{Camps} & \textcolor{red}{Camps} & \textcolor{red}{Camps} & \textcolor{red}{Camps} \\
\rowcolor{gray!10}\textbf{Success rate (\%)} & \textcolor{red}{100} & \textcolor{red}{100} & \textcolor{red}{100} & \textcolor{red}{90} \\
\bottomrule
\end{tabular}
\end{table}

Table~\ref{tab:fbs_vs_gain} summarizes the result metrics of this FBS assessment across varying cell gain values. In our experiments, an effective rogue base station attack did not depend on a single absolute distance threshold, but rather on the ability of the attacker's eNodeB to imitate the legitimate PLMN, outcompete the legitimate cell, and \textit{achieve signal dominance}. In practice, this required operating the rogue eNodeB in close proximity to the target vehicle within our shielded setup. Across the tested DL cell-gain range from 0 to -15 dB, we observed 100\% and 90\% (for -15 dB) attach successes, indicating that the attack remained effective even when the rogue cell was not configured at the highest tested gain. Because these measurements were collected using USRP B210 radios in a Faraday-tent environment, they should be interpreted as setup-specific practical conditions rather than universal real-world distance or power thresholds. Attackers equipped with higher-power SDRs, directional antennas, or better placement may achieve signal dominance at greater distances than in our setup.

The average active control-plane exchange per attach attempt lasted approximately four seconds and remained consistent across gain levels, as it is not signal-dependent. The Time-to-Attach was slightly longer when the rogue cell operated at lower power, potentially due to the selection logic and thresholds along with the minor signal fluctuations that may have occurred. However, for cell gains of -5 dB and higher, the TCU connected to the rogue cell almost immediately. Because the rogue eNodeB did not possess the credentials required to provide legitimate service as a real operator would, all attachments concluded in an incomplete attach state. The TCU remained camped after engaging with the rogue cell, resulting in a DoS condition. Importantly, under these conditions, we did not observe recovery to the legitimate network during an observation window of approximately 15 minutes while the rogue eNodeB remained active. In additional tests, recovery was not observed during a further observation window of approximately 10 minutes after the rogue eNodeB ceased transmitting. Subsequently, connectivity was restored only after manually rebooting the vehicle. Accordingly, we did not measure a spontaneous reconnection time to the trusted network within the monitored intervals. Each configuration was tested over 10 independent trials, with a single failure observed only at the lowest tested gain setting (-15 dB), resulting in a 90\% attach success rate.

We further extended our experimentation by simulating inter-cell handover scenarios to evaluate how the Tesla TCU responds to adversarial mobility conditions, where the rogue cells were configured to mimic real cells. By dynamically adjusting the signal strength, we forced the vehicle to transition between them as it would during a standard mobility event. Critically, the Tesla TCU attempted to preserve its existing security and session context during these transitions and did not reinitiate authentication, expecting the rogue cell to continue the existing session. Because the rogue cell lacked the credentials required to provide legitimate service, the connection ceased with the vehicle remaining in a camped state and unable to access any real cell during the monitored interval.

\begin{figure}[!t]
     \centering
     \includegraphics[width=\columnwidth]{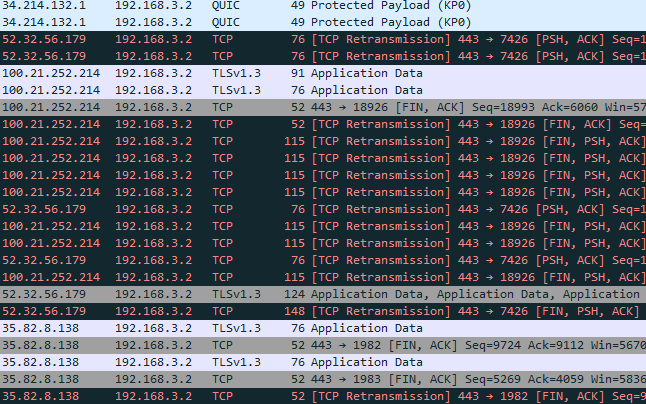}
     \caption{Tesla services failing during the FBS attacks.}
     \label{fig:service-degradation}
\end{figure}

Moreover, we noticed that by occupying the TCU in a bogus cell without uplink routing, attackers block the vehicle’s access to backend services responsible for software updates, remote unlock commands, real-time location tracking, and other reporting tasks. In more advanced cases, an attacker could also establish a MitM position using rogue UEs or downgrading to less secure protocols. Given that our focus is to evaluate the feasibility of FBS attacks, we intentionally refrained from executing more invasive subsequent attacks like a deeper payload alteration. Specifically, Figure~\ref{fig:service-degradation} shows the failure of Tesla's backend services during our FBS attacks in the data-plane. The depicted IP addresses are related to Amazon AWS that hosts Tesla applications such as ``maps-prd.go.tesla.services'' (IP 35.82.8.138), ``signaling-prd.vn.tesla.services'' (IP 52.32.56.179), and ``npuv-prd.usw2.vn.cloud.tesla.com'' (IP 100.21.252.214). Under typical conditions, the vehicle completes its connection with the network, before the attack takes place. Then, Figure~\ref{fig:lte-fbs-attacker} depicts a traffic example from the attacker's perspective, in which it records multiple Random Access and RRC attempts (several c-rnti identifiers) on the eNodeB side (srsRAN in this case) with the vehicle trying to establish its previous connection as well. Whereas, Figure~\ref{fig:lte-fbs-net} shows the network's perspective, in which the vehicle seems to disappear (i.e., radio connection lost) and cannot be reached with the paging process failing. Although full on-road testing was not conducted \textit{due to safety and legal constraints}, we empirically observed frequent backend communication during normal vehicle operation. Specifically, routine interactions (e.g., vehicle power cycling and driver-profile load, toggling lights/controls, and short low-speed movements within the tent) consistently triggered substantial bidirectional TCP traffic that was predominantly TLS-encrypted, limiting inspection to metadata such as timing, volume, and connection patterns. These observations indicate that backend connectivity is actively used during everyday operation and is therefore plausibly affected by adversarial conditions that disrupt cellular access.

The aforementioned findings highlight the risk of DoS against critical Tesla network functions, the potential for unauthorized attachment, and the opportunity for follow-up attacks such as LTE downgrades, session hijacking, or even MitM scenarios through rogue Network/UE integration. As a result, such threats could disrupt telemetry, remote commands, or safety-related services, posing significant security and availability risks for connected vehicles.

\noindent\textbf{$\bullet$ Fallback and Partial States Flaws.} To evaluate the resilience and failover behavior of Tesla’s telematics under adverse network conditions, we induced control- and data-plane failures while monitoring the TCU’s response via RRC, NAS, and IP traces. All tests used our previously validated custom network  with T-Mobile PLMN. Although these experiments use controlled network-side conditions to expose failure modes, the resulting behaviors are relevant because they determine how the TCU reacts after disruption by a wireless rogue-cell attacker.

Specifically, we began by simulating explicit control-plane rejections using standard NAS cause codes such as ``PLMN Not Allowed'', ``EPS Services Not Allowed'', ``Missing or Unknown APN'', ``MAC Failure'', and ``Unknown UE''. In response to these failures, the Tesla TCU consistently entered a repetitive reattachment loop, typically attempting the attach procedure five or more times in succession before halting or temporarily camping in an idle state. In no instance did the TCU revert to alternative connectivity options, such as switching to the eSIM profile. The reattachment behavior was deterministic and did not include randomized backoff or recovery strategies. To further evaluate fallback under data-plane degradation, we configured our core network to trigger PDN connectivity failures using standard bearer rejection causes such as ``Unknown PDN Type'', and ``PDN Type IPv4 Only Allowed''. In many of these cases, the vehicle completed the attach and remained in a partial-connectivity state, camping on the cell without receiving an IP address or completing data path setup. Despite the failure to establish end-to-end connectivity, the vehicle maintained the RRC connectivity and continued transmitting signaling messages, suggesting a lack of appropriate TCU failure detection.

Furthermore, we introduced network-level routing misconfigurations, including the removal of NAT rules, invalid routing table entries, and incomplete IP forwarding configurations. These setups allowed full RRC and NAS attachment, including successful PDN session establishment and bearer creation, while halting all uplink user-plane traffic. Under these conditions, the Tesla TCU continued transmitting application-layer traffic, such as HTTPS requests to Tesla backend servers, but received no responses. Despite the effective blackout of data communication, the TCU did not terminate the session, attempt reattachment to another network, or revert to WiFi as a fallback, but instead remained in this situation indefinitely.

Table~\ref{tab:failure-behavior} shows that across all scenarios, we observed no evidence that the TCU engaged in any form of fallback strategy, such as reverting to 3G, reconnecting using the eSIM profile, or preferring WiFi access, when LTE failed. Unlike modern smartphones, which support multiple active eSIM/SIM profiles, fallback logic, and more user-controlled options, Tesla’s implementation supports only a single active SIM at a time and lacks clear recovery mechanisms. This lack of dynamic switching suggests that the fallback logic between radio interfaces either misses proper health checks or is gated by higher-layer policies. As a result, this represents a critical limitation, by exposing the TCU to DoS conditions and prolonged network unavailability.

\begin{table}[!t]
\centering
\caption{Summarized TCU behaviors across failures. Here, \textcolor{green!60!black}{\ding{51}} indicates successful completion of the stage, \textcolor{red}{\ding{55}} indicates explicit failure at that stage, and `–-' indicates that the stage was not reached because a prior stage had already failed. The `\textcolor{red}{Yes}' and `\textcolor{red}{No}' indicate a violation.}
\label{tab:failure-behavior}
\renewcommand{\arraystretch}{1}
\setlength{\tabcolsep}{5pt}
\begin{tabular}{p{3.6cm}ccc}
\toprule
\textbf{Connection Stage} & \multicolumn{3}{c}{\textbf{Failures / Errors}} \\
\cmidrule(lr){2-4} & \textbf{Control} & \textbf{PDN} & \textbf{Routing} \\
\midrule
\midrule
\rowcolor{gray!10} Stage~1: Attach success & \textcolor{red}{\ding{55}} & \textcolor{green!60!black}{\ding{51}} & \textcolor{green!60!black}{\ding{51}} \\
Stage~2: PDN session & -- & \textcolor{red}{\ding{55}} & \textcolor{green!60!black}{\ding{51}} \\
\rowcolor{gray!10} Stage~3: Data connectivity & -- & -- & \textcolor{red}{\ding{55}} \\
\midrule
\multicolumn{4}{c}{\textbf{Behavioral outcomes}} \\
\midrule
\rowcolor{gray!10} Fallback triggered & \textcolor{red}{No} & \textcolor{red}{No} & \textcolor{red}{No} \\
Loop trapped & \textcolor{red}{Yes} & \textcolor{red}{Yes} & \textcolor{green!60!black}{No} \\
\rowcolor{gray!10} State transition failure & \textcolor{red}{Yes} & \textcolor{red}{Yes} & \textcolor{red}{Yes} \\
\bottomrule
\end{tabular}
\end{table}

\noindent\textbf{$\bullet$ Whitelisting and PLMN Issues.} To assess Tesla’s enforcement of operator whitelisting and carrier policies, we configured our LTE core with a wide range of MCC/MNC combinations, including legitimate, test, and arbitrary operator codes. Our objective was to determine whether Tesla’s TCU enforces a strict whitelist of allowed PLMNs during the LTE attachment process, and at what stage filtering actually occurs. PLMN allowability and selection are explicitly specified in 3GPP as a defense layer (including EHPLMN/VPLMN selection and forbidden-PLMN handling)~\cite{3gpp.23.122}. Therefore, vendor enforcement choices can materially change which networks the device will attempt to access and when it aborts the procedure. Consequently, overly permissive policies (or late enforcement) can increase the possibility of a UE engaging with adversarial cells, raising both attack surface and risk (We adhere to our A1/A2 threat model from Section~\ref{sec:risks}, and do not assume a privileged attacker).

To rigorously explore the behavior of Tesla’s TCU under different network identities, we implemented a PLMN (semi-automatic) testing methodology. This process iteratively configures the core network and radio stack with varying MCC/MNC combinations (and other related parameters), inserts corresponding SIM profiles, and monitors the TCU’s response at each stage of the LTE attachment and data service procedures. Algorithm~\ref{alg:plmn-brute} outlines the logic of our testing framework, which evaluates each PLMN configuration for attachment success, PDN session establishment, and final data service activation. Through this process, we discovered a multi-stage connectivity flow in Tesla’s LTE stack.

Once a physical SIM card is inserted, whether commercial or custom, the TCU immediately validates the SIM and IMEI (two distinct steps) and proceeds to perform network and data registration. These states can be reflected in the Tesla UI as ``Registered'' and ``Attached'' if successful, respectively. Although IMEI is independent of the SIM, inserting a SIM triggers a modem reinitialization/provisioning sequence (e.g., reconfiguring modem state prior to attach), during which the vehicle queries the cellular modem’s device identity. Tesla exposes this as a separate ``Valid IMEI” status in ``Service Mode''. Upon successful attachment, PDN session establishment and cell connectivity, the TCU attempts to negotiate roaming permissions to enable application-layer communication. In the final step, the ``Cell Link State'' is marked as ``online'' and is dependent on the successful establishment of the full connection. With the roaming approved, the link becomes fully operational for telemetry and user services. The results reveal a two-stage whitelisting process within the Tesla telematics unit; the first occurring during initial network selection and attachment, and the second during the activation of data network services (commonly referred to as ``roaming markets''). This layered behavior was confirmed through real-time analysis using Tesla's service mode interface, cellular connection logs, and the Tesla Toolbox software.

At the network selection stage, the vehicle accepted all physical SIM cards, regardless of operator identity, PLMN configuration, or authentication profile, without triggering any security warnings or restrictions. More importantly, it successfully completed both the LTE attach procedure and PDN session establishment across a variety of PLMNs, including standard commercial identifiers (e.g., 310260, 310150) and non-commercial or test identifiers (e.g., 00101, 00110, 99970). This indicates a permissive policy at the radio and core network levels. In contrast, modern smartphone platforms (e.g., later versions of iOS and Android) typically enforce stricter PLMN filtering, rejecting SIMs with non-authorized or test network identifiers by default. Additionally, Tesla's relaxed network policy can be modified dynamically using the Toolbox service, suggesting that PLMN validation is neither hardcoded nor enforced.

At the second stage, after attachment and PDN setup, the vehicle attempts to activate data network services (e.g., telemetry, internet access). The TCU restricts connectivity to a limited set of vendor-approved operators, likely based on backend validation or predefined roaming agreements. However, this filtering is implemented too late in the connection process to provide meaningful security. An attacker emulating a legitimate PLMN can bypass this constraint entirely. Even without full application-layer activation, the rogue network can still establish persistent control-plane connectivity with the TCU, exposing the vehicle to various attacks. Thus, the secondary filter functions more as a service-enablement gate than a protective control, and with significant limitations.

This design creates a risky security posture due to its permissive nature and complexity; while the vehicle may eventually deny certain forms of backend communication, it could willingly interact with potentially untrusted LTE infrastructure at the protocol level. Attacks such as IMSI catching, false base station exploitation, and even MitM positioning can still succeed, even with the current whitelisting policies in place. Overall, we confirm that Tesla’s whitelisting logic needs to be properly scoped and strictly enforced.

\noindent\textbf{$\bullet$ Control-Plane and Parameter Misconfigurations.} To evaluate the vehicle's security at the signaling level, we established full LTE connectivity and captured all control-plane exchanges during the attach, authentication, and PDN setup procedures. Using detailed logging of RRC and NAS messages, we monitored the interaction between the TCU and the network under various configurations and signaling profiles. However, rather than performing extensive protocol modifications or fuzzing, we adopted a targeted and non-invasive approach. Specifically, we altered core security parameters observing for any anomalies, misconfigurations or non-compliance with 3GPP. Such anomalies and misconfigurations can be used by a wireless attacker targeting the RRC and NAS explicitly.

Initial tests showed that Tesla’s attach procedures and authentication sequences align closely with those of commercial mobile devices. No irregularities were found in the attach flow ordering, or bearer management under standard configurations. The TCU consistently completed the EPS attach process, performed NAS authentication, and negotiated default bearers using expected message structures and standard cause values, even though repetitions of messages were not uncommon. We did not observe any Tesla-specific signaling extensions or proprietary messages at the RRC or NAS level during attach or bearer negotiation. All message types and procedures observed were consistent with standard 3GPP behavior. However, it remains unclear whether deeper protocol-level methods, such as fuzzing, could reveal implementation flaws. 

\begin{figure}[!t]
    \centering
    \begin{subfigure}{\columnwidth}
        \centering
        \includegraphics[width=\linewidth]{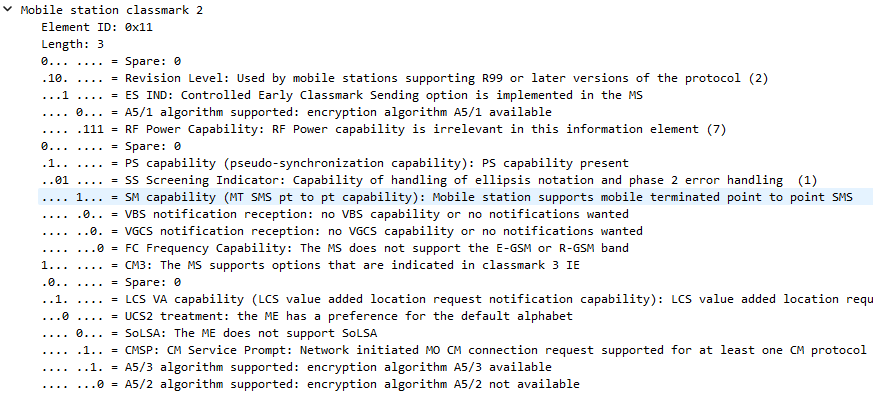}
        \caption{Mobile Station Classmark 2.}
        \label{fig:gsm-1}
    \end{subfigure}

    \vspace{0.0em}

    \begin{subfigure}{\columnwidth}
        \centering
        \includegraphics[width=\linewidth]{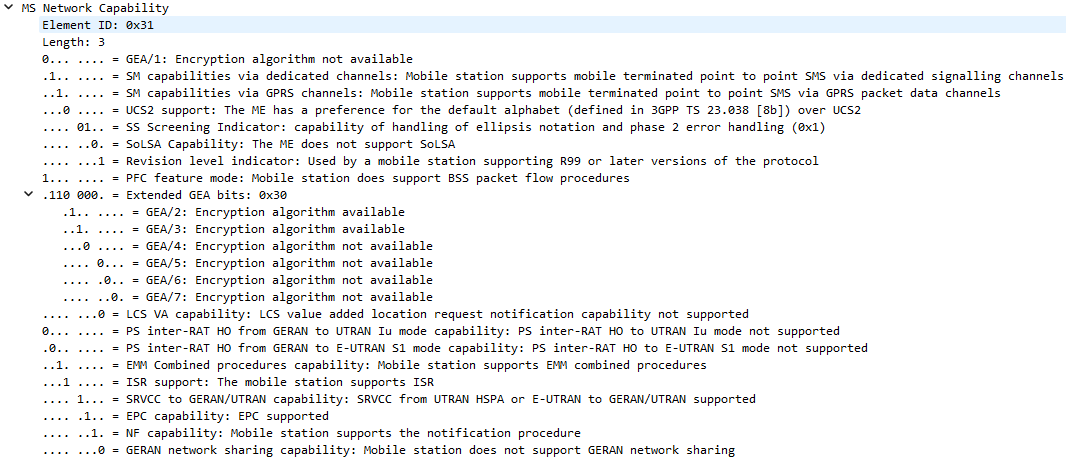}
        \caption{MS Network Capability.}
        \label{fig:gsm-2}
    \end{subfigure}

    \caption{Legacy support in Tesla TCU.}
    \label{fig:legacy-support}
\end{figure}

Unfortunately, our analysis also uncovered extensive legacy system support, particularly for 2G/GSM-based services and their associated cipher suites. The \texttt{Mobile Station Classmark 2} and \texttt{MS Network Capability} fields revealed the support of encryption algorithms such as A5/3 for GSM and GEA2/3 for GPRS data services, which are considered cryptographically outdated by modern standards~\cite{Dunkelman10:A5/3, Beierle21:GEA}. Moreover, the TCU advertises full compatibility with mobile-terminated point-to-point SMS over both dedicated, SRVCC to GERAN, and packet-switched GPRS signaling channels. This degree of backward compatibility, while potentially beneficial for roaming and network interoperability, significantly broadens the vehicle's attack surface and amplifies the vehicle's exposure to downgrade attacks. This design is worrisome since mobile vendors have started eliminating legacy features to upgrade security, as with the prominent example of Google's 2G Android support~\cite{GoogleSecurity2023, GoogleSecurity2024, HackerNews2024} and Apple's lockdown mode~\cite{apple_support_lockdown}. The presence of these capabilities, particularly in a vehicle environment lacking user-facing awareness, is considered a security liability. Figure~\ref{fig:legacy-support} shows an example of legacy capabilities from a captured NAS \texttt{AttachRequest}.

Additionally, the TCU declares support for several encryption and integrity protection algorithms. Notably, it supports the use of null ciphering (\texttt{EEA0}), alongside stronger algorithms such as SNOW3G (\texttt{128-EEA1}), AES (\texttt{128-EEA2}), and ZUC (\texttt{128-EEA3}). While null ciphering is permitted, it should not be prioritized during standard network negotiation. Importantly, it does not support null integrity (\texttt{EIA0}), indicating that integrity protection is correctly and consistently enforced, with preferred support for \texttt{128-EIA1/2/3}. Null integrity was rejected with a NAS \texttt{SecurityModeReject}. While our results suggest that the TCU behaves as a compliant LTE device under nominal conditions, its support for legacy protocols and weak ciphering modes, and the lack of transparency into its internal state significantly widen its attack surface.

\begin{figure}[!t]
    \centering
    \begin{subfigure}{\columnwidth}
        \centering
        \includegraphics[width=0.85\linewidth]{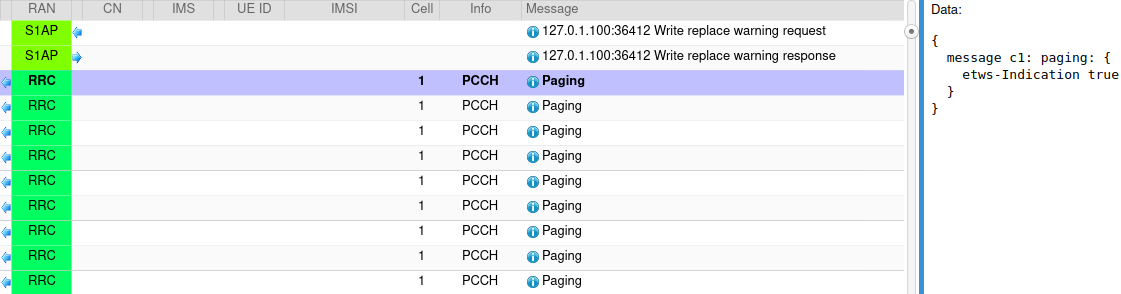}
        \caption{ETWS warnings.}
        \label{fig:warmings}
    \end{subfigure}

    \vspace{0.0em}

    \begin{subfigure}{\columnwidth}
        \centering
        \includegraphics[width=0.85\linewidth]{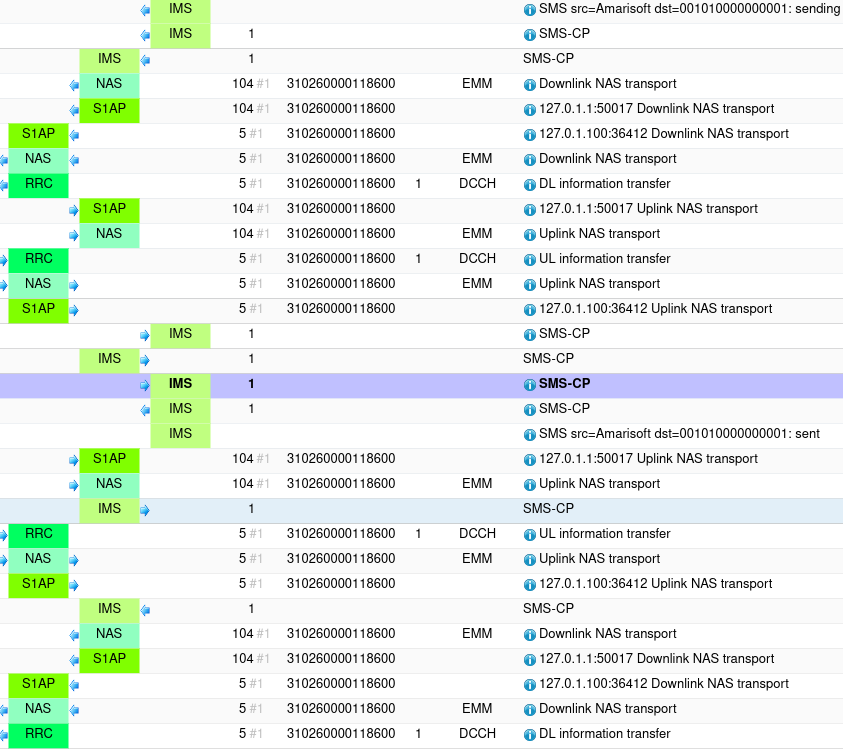}
        \caption{SMS process via IMS.}
        \label{fig:sms}
    \end{subfigure}
    
    \caption{Warning broadcasts and SMS message examples.}
    \label{fig:tesla-sms-warning}
\end{figure}

\noindent\textbf{$\bullet$ SMS and Emergency Issues.} To evaluate how the Tesla TCU handles broadcast alerts and SMS messages, we performed controlled injection tests targeting both emergency notification protocols and standard SMS transport mechanisms. Our testing covered two categories: (1) Emergency alerts via ETWS (Earthquake and Tsunami Warning System) and CMAS (Commercial Mobile Alert System), and (2) direct SMS injection using IMS delivery. Emergency broadcast tests were run against both the eSIM and physical SIM. These tests do not assume a separate insider attacker, rather they assess whether additional wireless-facing interfaces are exposed that a wireless attacker may be able to trigger or misuse. Figure~\ref{fig:warmings} and~\ref{fig:sms} illustrate traffic examples from our testing.

For emergency alerts, we broadcast both normal and spoofed ETWS/CMAS messages from our eNodeB while the vehicle was RRC-connected and fully attached to the network. In all cases, we confirmed that the messages were successfully transmitted by the base station. However, the vehicle did not display any user-facing alert or audible warning, regardless of the SIM type or alert category. Based on our black-box evaluation, we cannot determine whether these messages were filtered, silently processed, ignored by design, or handled elsewhere in the software stack. While this behavior may be intentional, the absence of any observable user-facing alert warrants further investigation. In particular, if legitimate emergency broadcasts are not surfaced to vehicle occupants under operational conditions, passengers may need to rely on external devices, such as smartphones, to receive time-sensitive safety notifications.

In the SMS-injection experiments, we transmitted several well-formed, text-based messages to the TCU over the LTE signaling. An attacker could deliver such messages through standard carrier channels even as an ordinary subscriber (or SMS-sending service) if the vehicle’s MSISDN (Mobile Station ISDN Number) is known/discoverable, and under conditional legacy-fallback settings (e.g., 2G downgrade) with rogue cells. In all cases, the vehicle received the messages validating its open interface, confirmed delivery at the protocol level, and remained RRC-connected for the duration of the exchange. However, the Tesla interface did not display any visual indication that a message had been received. This behavior is consistent with a closed SMS stack in which the TCU handles incoming messages internally but does not expose them to the user or offer UI-level access. From a security perspective, this architecture introduces silent processing paths that are invisible to the end-user, and therefore, non-trivial to audit or monitor. Unlike smartphones, where users can observe incoming SMS messages, deploy message-filtering applications, or detect anomalies in real time, the TCU operates as a black box. Our experiments establish that the SMS interface remains reachable and largely opaque to the end user, but do not evaluate the handling of malformed, binary, or silent SMS payloads. However, as discussed in Section~4.2, prior work on other platforms has shown that insufficiently protected SMS-processing paths may expose systems to risks such as denial of service or code execution. Accordingly, the security properties and operational purpose of the exposed SMS interface warrant further investigation. If SMS support is not operationally required, stricter filtering, logging, or disablement should be considered.

\section{Discussion and Broader Implications}

\noindent\textbf{Regulatory Implications.} Our findings have direct implications for regulatory compliance with UN Regulations 155 (Cyber Security)~\cite{UNECE-R155} and 156 (OTA Updates)~\cite{UNECE-R156} mandatory for vehicle type approval in many jurisdictions. R155 requires secure communication channels, resilience against DoS attacks, protection of identity data, and capabilities for detection, response, and forensics. R156 mandates the secure, reliable, and auditable software updates.

Our experiments reveal several violations of these principles. Tesla’s telematics stack is vulnerable to IMSI catching and rogue base station attachment due to weak design and implementation practices and standards, raising concerns under R155's provisions for communication security (Annex 5, 7.3.3--7.3.6). The persistence of advertised weak/legacy cryptographic algorithms (e.g., null ciphering) and silent SMS/emergency alert reception implicate expectations for robust cryptography and attack detection (Table A1, M10–M13). Additionally, our FBS-based DoS attacks block access to backend update services, undermining the availability and integrity of OTA delivery in violation of R156 Clause 7.2.1.1. The absence of any user-visible alert about these failures raises concerns under Clause 7.2.2.2, which mandates transparency and user awareness during update processes. Table~\ref{tab:r155-r156-mapping} provides further details on the mapping of each vulnerability/attack to a relevant UN requirement. ISO certifications, such as 21434~\cite{ISO-SAE-21434} and 26262~\cite{ISO26262_2018}, may also be impacted, as weaknesses in PLMN whitelisting, connectivity disruptions, or spoofed signals undermine 21434's emphasis on systematic threat analysis and risk assessment, and interfere with 26262's requirements for functional safety (e.g., emergency updates). While not always direct non-conformance, such issues highlight residual risks in conflict with these standards. Together, the findings show that known mobile-layer threats can create systemic regulatory concerns.

\noindent\textbf{Systemic Risks \& Attacker Requirements.} Our experiments demonstrate that even high-end commercial vehicles with advanced connectivity are susceptible to fundamental LTE-layer attacks. These vulnerabilities carry significant risks, particularly because cellular connectivity underpins remote operations and safety-relevant services (e.g., emergency assistance and update delivery), even if core driving safety functions may not depend on continuous cellular connectivity. However, we also acknowledge that in severe cases, attacks may degrade emergency-assistance workflows and increase risk in already-dangerous situations, potentially exacerbating outcomes such as property damage or injury. 

While the attacks described can be executed using commercially available SDR hardware and open-source LTE stacks (e.g., srsRAN, Amarisoft), real-world adversaries may face several constraints. These include the need for close proximity to the vehicle, careful transmission power tuning to overpower legitimate base stations without detection, sufficient coverage for mobile targets, and hardware capable of low-latency, precise interaction with the LTE protocol stack. Additionally, vehicles in motion reduce the attacker's dwell time, posing challenges for sustained exploitation. However, determined and advanced attackers may overcome these limitations. For instance, targeting vehicles in static environments, such as parking garages, service centers, or border crossings, mitigates mobility constraints and enables prolonged exposure. Similarly, directional antennas (highly focused RF transmission improves signal-to-noise ratio), mobility (e.g., drone-mounted equipment), and multiple rogue gNodeBs  (e.g., roadside) can help maintain proximity, signal dominance and coverage.

\noindent\textbf{Cellular Connectivity in Other Automobiles.} During our experimentation, we did not identify any behavioral differences between the tested vehicles. While Tesla’s hardware and software configurations may vary across markets, vehicle generations, and regional carrier agreements, it is reasonable to assume that most mainstream Tesla models (i.e., Model 3, Y, S, X, and Cybertruck) rely on embedded modems for cellular connectivity. However, not all models support physical SIM cards; many are restricted to pre-provisioned eSIM profiles. We also examined closely the connectivity practices of other major manufacturers via access to various vehicles, such as BMW, Volvo, Toyota, Mercedes, Lexus, Ford, and Hyundai, which rely on factory-installed eSIMs.

A key challenge for independent security evaluation is that many vehicles provide \emph{little to no researcher-controlled access} to the cellular subscription and capabilities. Unlike smartphones, where researchers can swap SIMs/eSIM profiles, inspect network selection settings, and examine baseband behavior (e.g., Qualcomm's QXDM), automotive TCUs are frequently provisioned with \emph{factory-installed, non-user-configurable eSIM profiles}. In these systems, the cellular identity and configurations (e.g., carrier policies, roaming configuration, APN provisioning) are locked behind OEM/carrier provisioning, without support for custom eSIM profiles or controlled SIM swapping. Additionally, vehicles typically expose only coarse ``connected/not connected” indicators rather than cellular-layer diagnostics (e.g., connection failure reasons), which further constrains systematic measurement. As a result, researchers cannot delve into the details of internal cellular processes and protocols with over-the-air testing, apart from limited pre-authentication signaling (at best) that reveals very little about the vehicle's holistic security posture. So, we encourage OEMs to improve the transparency and testability.

Despite these access limitations, several of the behaviors we observe may \textit{extend beyond Tesla}, particularly due to the common adoption of similar embedded modems, telematics integration patterns, and LTE/5G procedures across OEMs. Some observations, such as exposure to pre-authentication signaling, IMSI disclosure, and susceptibility to rogue-cell selection, arise from limitations of cellular procedures and are not Tesla-specific (\textit{cellular-design-level}). Other observations, including advertised legacy capabilities, PLMN handling, fallback and recovery behavior, and user-facing visibility of SMS or emergency messages, may depend on modem capabilities, TCU integration, or OEM telematics policies (\textit{implementation-dependent}). Because our evaluation is black-box, we report the externally observable behavior of the tested Tesla platforms and do not assign each observation conclusively to a specific component. We therefore treat Tesla as a case study and provide a methodology intended to support comparable evaluations of other vehicles when suitable access is available.

\noindent\textbf{Applicability to 5G SA/NSA.} Although our experiments focus on LTE, several of the threat classes we study (and related ones) are expected to remain relevant in 5G deployments~\cite{eleftherakis25:sok, Shijie25:sni5gect, Bitsikas21:Handovers}. While 5G introduces meaningful security improvements, these do not eliminate fundamental and implementation-based risks. For example, although 5G conceals the permanent subscriber identifier via SUCI, prior work shows that SUCI-based mechanisms can still enable UE profiling or linkability~\cite{Chlosta21:suci-catchers}. We therefore expect our methodology to be adaptable to 5G-connected vehicles, and suggest systematic evaluation of 5G-capable TCUs across 5G deployments.

\section{Mitigation Strategies}

We argue that mitigations should represent a security baseline that should be integrated across all connected vehicles (not Tesla-specific). However, implementing and validating these mitigations demands low-level access to each vehicle’s TCU and modem, which is typically \textit{restricted to manufacturers}.

\noindent\textbf{IMSI Catching.} To address this, the vehicle should suppress permanent identifier transmission unless mutual authentication with a trusted PLMN is confirmed. It should not respond to any messages on networks that fail validation checks during early radio procedures. Support for randomized GUTI reallocation and identity protection mechanisms could be incorporated. Encryption-based obfuscation and strict timer-based reallocation of temporary identifiers can help against tracking too. However, since radical design modifications are difficult to adopt, we believe it is reasonable for the vehicles to upgrade and leverage by default a 5G Standalone connectivity that provides stronger security guarantees through the support of SUCI~\cite{3gpp.23.003}, though 5G coverage limitations could pose a problem. In such a scenario, LTE can serve as a fallback.

\noindent\textbf{False Base Stations.} To ideally prevent this, the vehicle and the network should mutually enforce the use of signed broadcast messages (e.g., MIB, SIB1) and the validation thereof. Monitoring and logging malicious activity can also assist in identifying threats. These could ensure that only verified operators are accepted before any connection takes place. Additionally, handover events to unknown cells should be blocked unless cryptographically validated. Ultimately, 3GPP 33.809~\cite{3gpp.33.809} provides robust measures against FBS attacks, however, it is unknown whether such measures will ever be deployed in real systems. The performance demands, mass adoption effort and the extensive design modifications may render the defenses impractical in real-world scenarios. 

\noindent\textbf{Whitelisting and PLMN Filtering.} We argue for a simpler and stricter whitelisting methodology that includes only legitimate and authorized network operators. Network selection filtering should only initiate NAS procedures when the cell identity matches a pre-validated set of legitimate MCC/MNC combinations. The TCU must prevent attachment to test PLMNs or unknown MNCs unless explicitly provisioned by a secure profile. Dynamic management of roaming partners is necessary to revoke trust relationships.

\noindent\textbf{Fallback and Partial State Handling.} In this case, attach retries could follow exponential backoff with randomized timers. The TCU should detect persistent failures and switch to an eSIM profile, PLMN, or WiFi network. End-to-end health monitoring, such as failure to reach Tesla backend services, can trigger fallback logic. The ability to dynamically switch access interfaces (e.g., from LTE to WiFi) is essential in mobile environments. This indicates that there should be an implementation to perform the dynamic switching and coordinate radio technologies accordingly. In addition, more information should be provided to the user via the control screen, who may assist the vehicle in the fallback procedures. However, such measures necessitate additional implementation that may also increase the complexity of vehicle operations.

\noindent\textbf{Control-Plane and Capability Vulnerabilities.} Despite mostly standard behavior, the TCU advertises support for 2G/GSM services, weak encryption algorithms (e.g., A5/3, GEA3) and null ciphering. While this may improve backward compatibility, it amplifies exposure to cellular attacks. These ciphers should be disabled and replaced with robust algorithms (e.g., AES' \texttt{EEA2} and \texttt{EIA2}) by default. Additionally, GSM services should be completely disabled, and the protocol stack logic should strictly validate NAS and RRC message structures and reject malformed fields to reduce susceptibility to malicious control-plane message injections.

\noindent\textbf{SMS and Emergency Message Handling.} Our tests showed protocol-level SMS reception without user-visible indications and no visible or audible alert following ETWS/CMAS transmission under the tested conditions. Because the internal handling path remains unclear, OEMs should assess the operational need for these interfaces, apply appropriate filtering and logging, and surface legitimate alerts to users (i.e., more transparency for users). If SMS support is unnecessary, disabling the interface should be considered.

\section{Conclusion}

This paper presents a practical Tesla-based case study of LTE connectivity security in connected vehicles under adversarial conditions. By integrating production Tesla vehicles into a controlled custom LTE infrastructure, we established a reproducible methodology for studying LTE security behaviors in a production telematics platform under controlled adversarial conditions. Our findings expose significant security gaps in the evaluated Tesla platforms and illustrate how connected-vehicle telematics can remain exposed to known cellular threats. Our case-study- and real-world-driven approach provides a practical foundation for future evaluations and improvements in cellular security across connected-vehicle platforms, as adoption of cellular connectivity increases.

\section*{Acknowledgement}

The authors gratefully acknowledge the support of NSF Grant 2144914, Google PhD Fellowship, and the United States Coast Guard Academy. The views expressed in this publication are those of the authors and do not necessarily represent the views of the United States, the Department of Homeland Security, or the United States Coast Guard. Additionally, we gratefully acknowledge Consumer Reports for providing the test vehicles and facilities that made this security analysis possible.

\bibliographystyle{plain}
\bibliography{references}

\appendix

\begin{table*}[!t]
\centering
\small
  \setlength{\tabcolsep}{5pt}
  \renewcommand{\arraystretch}{1.05}
\begin{tabularx}{\textwidth}{
      >{\raggedright\arraybackslash}p{3.0cm}
      >{\raggedright\arraybackslash}p{5.5cm}
      >{\raggedright\arraybackslash}X}
\toprule
\rowcolor{gray!25} \textbf{Identified Vulnerability} & \textbf{Regulatory Requirements (R155 / R156)} & \textbf{Mapping Justification} \\
\midrule
IMSI catching \& identity exposure & R155 Annex 5 Table B1, Threat 7.1 (interception/monitoring), Mitigation M12; Vehicle-type 7.3.3, 7.3.7 & Confidential identifiers must be protected; vehicle fails to mitigate interception or to provide detection/forensics. \\
\hline
\rowcolor{gray!10} Rogue base station \& lax PLMN enforcement & R155 Annex 5 Table B1, Threat 6.1 (accepting untrusted info), 6.2 (MitM), Mitigation M10; Threat 8.1 (DoS), Mitigation M13; Vehicle-type 7.3.3–7.3.7 & OEM must authenticate sources, prevent spoofing, and detect DoS; Tesla attaches to untrusted PLMNs without recovery. \\
\hline
Insecure fallback \& attach loops & R155 Table B1 Threat 8.1 (DoS), Mitigation M13; Vehicle-type 7.3.4, 7.3.6, 7.3.7. R156 7.1.3.1–7.1.3.3, 7.2.2.1.1, 7.2.2.2, 7.2.2.4, 7.2.2.5 & Persistent attach loops show lack of DoS recovery (R155) and violate R156 requirements for secure, recoverable, and user-notified OTA processes. \\
\hline
\rowcolor{gray!10} Legacy \& weak crypto (e.g., null cipher, A5/3) & R155 Table B6 Threats 26.2/26.3, Mitigation M23; Vehicle-type 7.3.8 & Legacy and null-cipher capabilities should be carefully restricted in safety-relevant communications; the tested TCU still advertises support for such capabilities. \\
\hline
Protocol-level SMS reception without user-visible indication & R155 Table B1 Threat 6.1, Mitigation M10; Table B4 Threat 16.2 (telematics manipulation), Mitigation M20; Vehicle-type 7.3.7 & External messages should be authenticated and monitored; the reachable and user-opaque SMS interface creates an attack surface that warrants further investigation. \\
\hline
\rowcolor{gray!10} No user-visible alert following emergency broadcasts & R155 Table B1 Threat 6.1, Mitigation M10; Vehicle-type 7.3.6, 7.3.7 & Emergency alerts from untrusted sources should be validated; spoofed ETWS/CMAS transmissions did not produce a visible or audible alert under the tested conditions, while internal handling remains unknown. \\
\hline
Services blocked by rogue LTE base station & R155 Table B2 Threat 13.1 (DoS vs. update rollout), Mitigation M3. R156 7.1.3.1 – 7.1.3.2, 7.2.2.1.1, 7.2.2.2, 7.2.2.4, 7.2.2.5 & Disrupting access to update services is a recognized DoS threat (R155) and may affect OTA availability, recovery, and user-awareness requirements under R156. \\
\hline
\rowcolor{gray!10} Lack of user awareness of update failures & R156 7.2.2.2 (inform before), 7.2.2.4 (inform after) & Users should receive relevant connectivity and update-status information; Tesla provides no user-visible feedback during connectivity disruption. \\
\hline
Supplier-dependent stack & R155 Vehicle-type 7.2.2.5 (supplier dependencies), 7.3.2 (supplier-related risks) & R155 requires managing supplier risks; risks involving shared cellular components or integration patterns may propagate across OEMs, not just Tesla. \\
\bottomrule
\end{tabularx}
\caption{Mapping of identified Tesla LTE vulnerabilities to relevant UN R155 and R156 requirements, with justification.}
\label{tab:r155-r156-mapping}
\end{table*}

\section{LTE Architecture} \label{sec:architecture}

\begin{figure}[!t]
     \centering
     \includegraphics[width=\columnwidth]{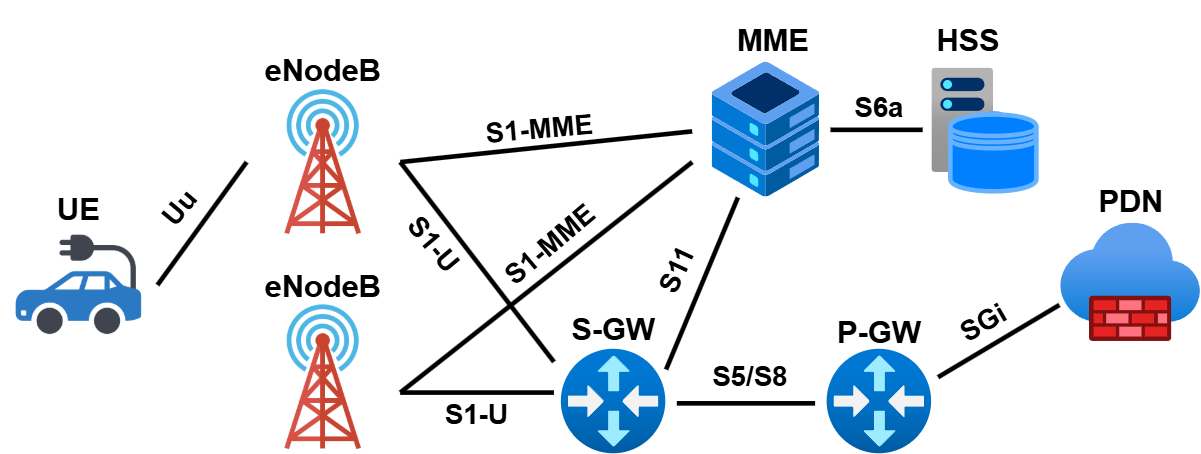}
     \caption{The architecture of an LTE network.}
     \label{fig:lte-network}
\end{figure}

Long-Term Evolution (LTE)~\cite{3gpp.23.401} is designed around a simplified, packet-centric framework known as the Evolved Packet System (EPS), which consists of the Evolved Universal Terrestrial Radio Access Network (E-UTRAN) and the Evolved Packet Core (EPC). The \textit{User Equipment (UE)} is the device that connects to the LTE network, which could be a smartphone, IoT module, or an automotive TCU. The E-UTRAN comprises the \textit{evolved Node B (eNodeB)} stations, which communicate with the UE and manage the radio connectivity, resource allocation, quality-of-service (QoS) policies, and mobility. The EPC is responsible for user data routing, mobility management, subscriber authentication, and gateway functions to external networks. It consists of the \textit{Mobility Management Entity (MME)}, which is a control-plane entity that coordinates the subscriber mobility, authentication, and session management, and the \textit{Home Subscriber Server (HSS)}, which is a centralized database holding user profiles, subscription information, and authentication credentials. The LTE network also contains specialized routing entities, the \textit{Serving Gateway (S-GW)}, which is a data-plane node anchoring user IP traffic during mobility events acting as an intermediary and maintains UE context, and the \textit{Packet Data Network Gateway (P-GW)}, which is the gateway providing connectivity to external networks, such as the internet or \textit{Packet Data Networks (PDN)}. Figure~\ref{fig:lte-network} illustrates the LTE architecture.

\begin{algorithm}[!t]
\caption{Whitelisting and roaming policy fingerprinting.}
\small
\label{alg:plmn-brute}
\begin{algorithmic}[1]
\ForAll{PLMN $\in$ candidate\_plmns}
    \State Configure eNodeB and EPC with PLMN parameters
    \State Power off Tesla vehicle network interface
    \State Restart LTE stack (eNodeB + EPC)
    \State Insert SIM programmed with same PLMN
    \State Power on vehicle and monitor TCU behavior
    \If{SIM Detected and Validated}
        \State Attempt for LTE Attach
        \If{Attach Successful}
            \State Attempt PDN Session Establishment
            \If{PDN Success}
                \State Verify Roaming \& Cell Link State
                \If{Application-Layer Active}
                    \State \textbf{Log:} PLMN $\rightarrow$ Full Connectivity
                \Else
                    \State \textbf{Log:} PLMN $\rightarrow$ PDN, No Roaming
                \EndIf
            \Else
                \State \textbf{Log:} PLMN $\rightarrow$ Attach Without PDN
            \EndIf
        \Else
            \State \textbf{Log:} PLMN $\rightarrow$ Rejected at Attach
        \EndIf
    \Else
        \State \textbf{Log:} PLMN $\rightarrow$ SIM Invalid
    \EndIf
\EndFor
\end{algorithmic}
\end{algorithm}

\begin{table}[!t]
\centering
\caption{Relation of findings to the wireless attacker model. A1 denotes a passive wireless observer, while A2 denotes an active attacker with rogue bases/transceivers. We use A1/A2 only where the finding itself can directly include both passive observation and active exploitation. A2-only rows may still involve passive reconnaissance in practice, but are mapped to the attacker class most directly responsible for the attack or its effect. The role column indicates whether the finding constitutes a direct attack, an enabling condition, or a system property that amplifies the effect of a wireless attack.}
\label{tab:threat_mapping}
\rowcolors{2}{gray!12}{white}
\begin{tabular}{p{0.42\columnwidth} p{0.13\columnwidth} p{0.33\columnwidth}}
\toprule
\rowcolor{gray!25}
\textbf{Finding} & \textbf{Attacker} & \textbf{Role} \\
\midrule
IMSI catching & A1/A2 & Direct attack vector / privacy exposure \\
False base station & A2 & Direct attack vector \\
Fallback / partial state & A2 & Impact/persistence amplifier \\
PLMN allowlisting & A2 & Attack enabler \\
Control-plane / capability handling & A2 & Attack-surface amplifier \\
SMS / emergency handling & A2 & Post-selection exposure \\
\bottomrule
\end{tabular}
\end{table}

\section{Experimental Equipment} \label{sec:equipment}

\subsection{Tesla Model 3} In this part we provide more information about the Tesla features.

\noindent\textbf{Modem.} The Tesla vehicle's telematics system is powered by the Quectel AG525R-GL modem. Designed for global LTE connectivity, it supports 4.5G LTE with Cat 6, 9, 12, and 16 capabilities depending on the region, offering peak download speeds of up to 600 Mbps (LTE Cat 12) and upload speeds of up to 150 Mbps. It features a wide range of frequency bands covering LTE-FDD, LTE-TDD, UMTS, CDMA (optional), and GSM with 30 Mbps Tx/Rx. For data processing and communication, it supports TCP, UDP, FTP(S), HTTP(S), TLS, MQTT, and QMI protocols, ensuring versatile connectivity. The modem interfaces with Tesla’s onboard systems via a PCIe bus, USB, RGMII, SPI, and multiple UARTs. It includes a single USIM slot (expandable to dual SIM) for network access.

\begin{figure}[!t]
     \centering
     \includegraphics[width=\columnwidth]{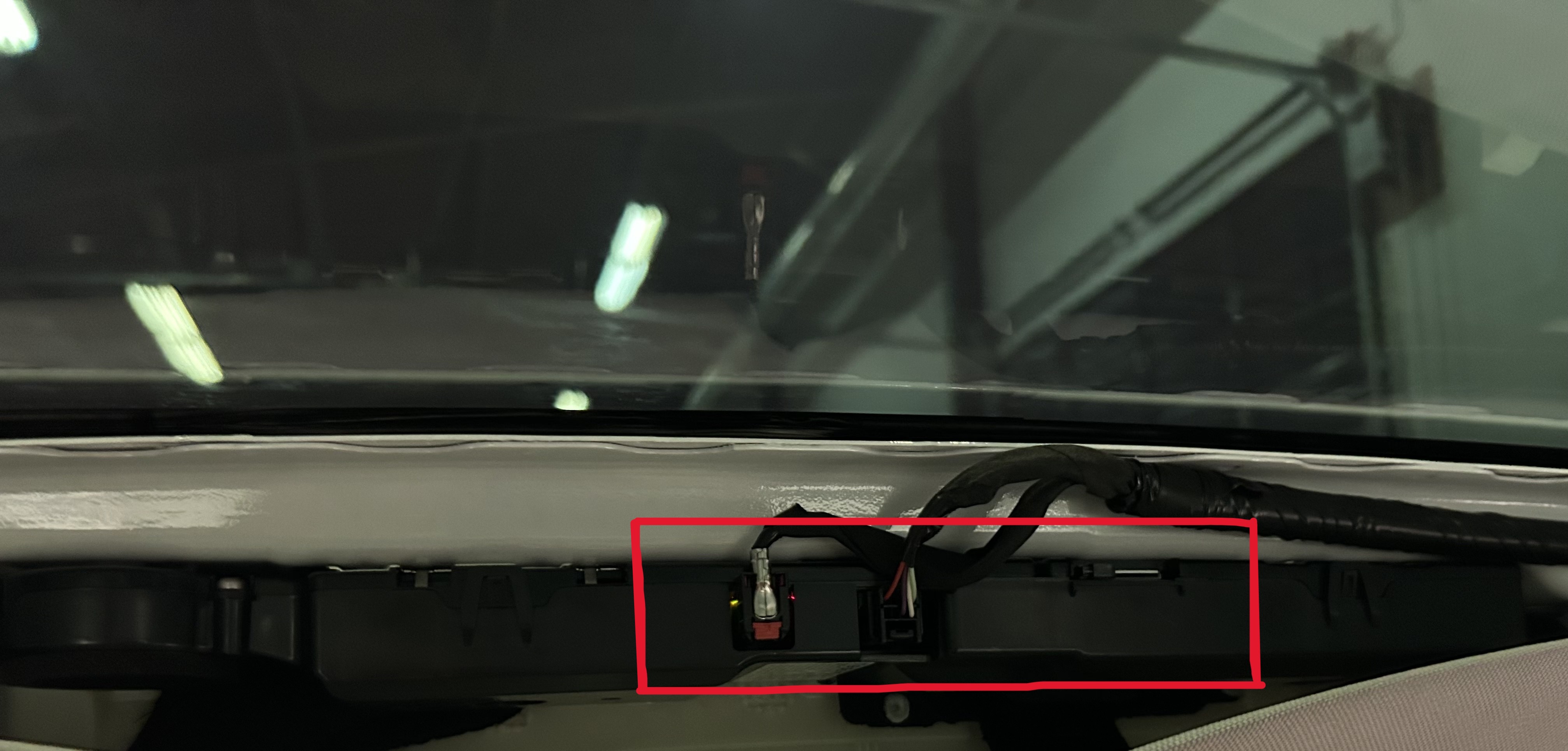}
     \caption{LTE modem and SIM slot.}
     \label{fig:lte-modem}
\end{figure}

\noindent\textbf{eSIM \& Physical SIM.} The internal eSIM is a T-Mobile (or AT\&T) profile that is enabled by default. The automobile uses that profile to connect to available commercial networks and to enable roaming for internet access. The physical SIM slot (see Figure~\ref{fig:lte-modem}) is secured by a protective cover that must be removed to expose the modem’s hardware interfaces. Once the cover is removed, users can physically insert a SIM card, though the system still prioritizes the default eSIM unless manually reconfigured via the Tesla Toolbox.

\noindent\textbf{Operating System.} The operating system environment of the TCU is a restricted Linux-based system (i.e., Linux infoz 5.4.284-INFOZ), tailored for automotive telematics and network management. It supports a limited set of commands, restricting user access to core functionalities while allowing controlled interactions for diagnostics, connectivity management, and updates. The restricted shell environment ensures security by preventing arbitrary modifications or unauthorized command execution. We utilized the commands in order to retrieve more information about the use of the LTE network and about the interaction with the modem.

\begin{figure}[!t]
     \centering
     \includegraphics[width=\columnwidth]{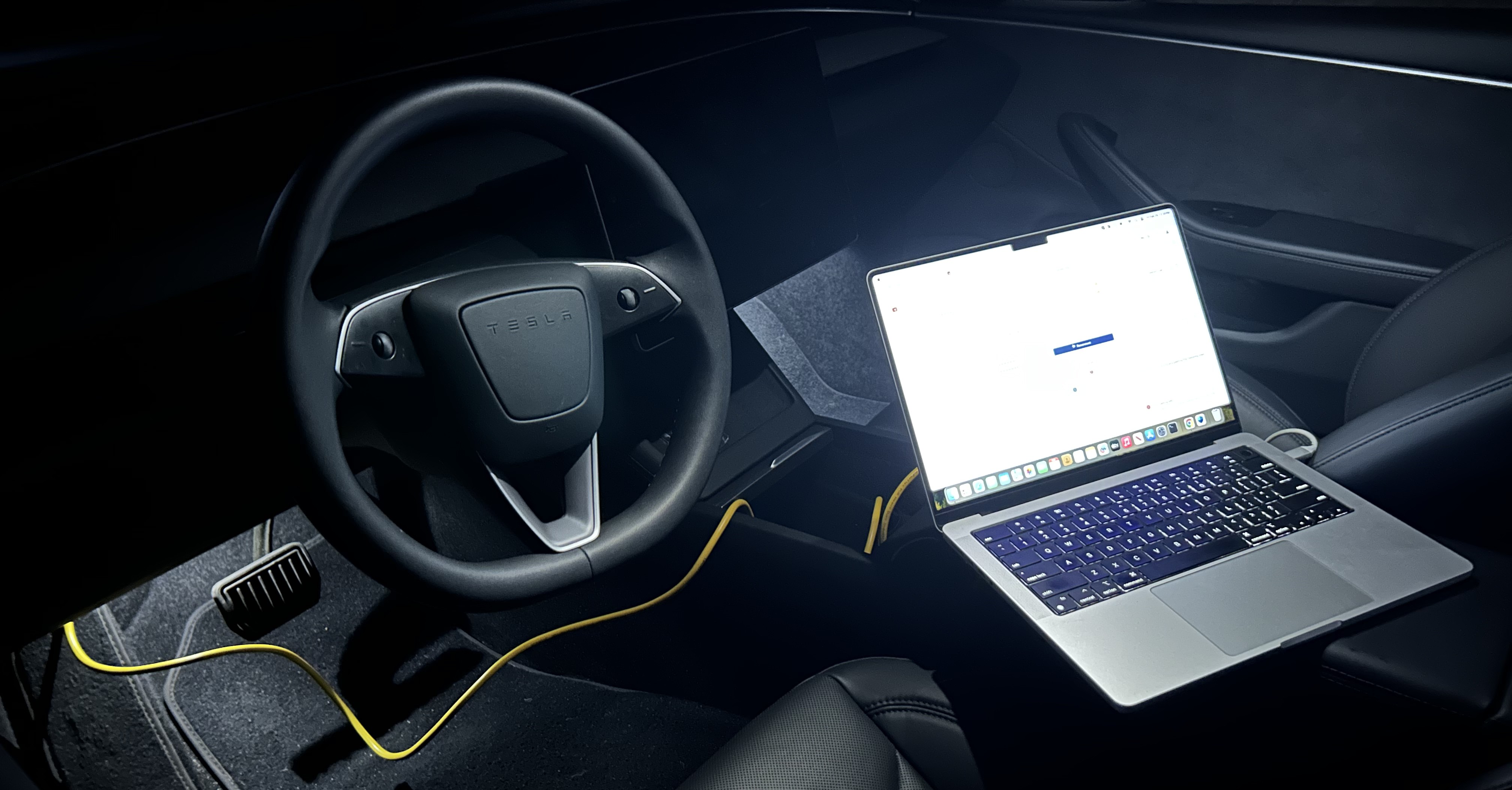}
     \caption{Tesla OBD port connected to the laptop.}
     \label{fig:lte-obd2}
\end{figure}

\noindent\textbf{OBD Port.} The OBD port in Tesla vehicles (see Figure~\ref{fig:lte-obd2}) is a proprietary diagnostic port using Ethernet for service and telemetry.  This port facilitates vehicle diagnostics, firmware updates, and telemetry logging, but access is restricted, requiring specific authorization with subscription to retrieve meaningful data. It permits access to the car's parametrization for cellular networks via the Tesla Toolbox software~\cite{TeslaToolbox}, which is a locally web-hosted tool.

\subsection{Faraday Tent} The Faraday tent used for testing provides a high level of RF attenuation, effectively isolating LTE signals from external commercial networks and preventing unintended interference with real carriers. The tent is designed to attenuate signals by 93 dB across the relevant LTE frequency bands, including 700 MHz (Band 12/13/14/17), 850 MHz (Band 5), 1800 MHz (Band 3), 1900 MHz (Band 2/25), and 2600 MHz (Band 7/41). This isolation ensures that the conducted tests remain contained within a controlled RF environment, eliminating unwanted external influences that could distort signal strength, interference levels, or device behavior. Typically, omnidirectional LTE antennas are positioned at a fixed height and orientation relative to the vehicle under test to ensure consistent signal propagation. The necessary connectivity cables (e.g., Ethernet) are inserted into the tent via its interface.

\subsection{Amarisoft and srsRAN} In our experimentation, we utilized both \textit{Amarisoft} and \textit{srsRAN} to create and manipulate LTE network environments for testing the Tesla’s connectivity. Amarisoft was configured to provide a fully functional LTE network for complete device attachment, authentication, and service provisioning, while srsRAN was primarily used for conducting IMSI catching and FBS attacks. Although IMSI catching was successfully performed using both platforms, we also targeted the legitimate T-Mobile eSIM with Amarisoft. The latest stable versions of both software suites available were used. For Amarisoft, we configured and utilized the MME, eNodeB, and IMS modules, whereas for srsRAN, we primarily employed the EPC and eNodeB components. Amarisoft was deployed on a dedicated desktop system equipped with multiple SDRs, whereas srsRAN was installed on a Lenovo ThinkPad i7 running Ubuntu and interfaced with an Ettus USRP B210.

\noindent\textbf{Configuration and Initialization.} Both software suites required specific configurations to establish an LTE network. In Amarisoft, configuration files such as \texttt{enb.cfg} for the eNodeB, \texttt{ims.cfg} for the IMS and \texttt{mme.cfg} for the MME were adjusted to match the required setup. Additionally, we made adjustment in the credential database to support the Tesla as a UE. To run Amarisoft, we simply executed:

\begin{verbatim}
$./lte-init.sh (routing setup)
\end{verbatim}

\begin{verbatim}
$./ltemme config/mme.cfg
\end{verbatim}

\begin{verbatim}
$./ltemme config/ims.cfg
\end{verbatim}

\begin{verbatim}
$./lteenb config/enb.cfg
\end{verbatim}

For srsRAN, we modified configuration files like the \texttt{enb.conf}, \texttt{rr.conf} and \texttt{epc.conf}, in order to define key network parameters, such as the \texttt{mcc}, \texttt{mnc}, \texttt{n\_prb} (number of physical resource blocks), and \texttt{dl\_earfcn} (downlink frequency). The eNodeB and EPC were simply started with \texttt{sudo srsenb}, and \texttt{sudo srsepc}.

\noindent\textbf{Base Station Setup and Monitoring.} Once initialized and during the experimentation process, both Amarisoft and srsRAN required active monitoring to track network performance and UE behavior. For example, in Amarisoft, the command \texttt{t} displayed trace metrics. For srsRAN, entering \texttt{t} during execution enabled similar tracing functionalities. Additional tools such as Wireshark~\cite{Wireshark} and tcpdump~\cite{tcpdump} were used to capture and analyze LTE and data-plane traffic. Other network tools/commands, such as Ping~\cite{ping}, iPerf~\cite{iperf}, and traceroute~\cite{traceroute}, were used for our network assessments and health checks.  

\section{Vehicle Network Integration Methodology} \label{sec:method}

Since the Tesla’s TCU operates as a closed system with limited direct configuration access, our methodology follows a progressive approach, beginning with observing normal connectivity behavior on commercial networks before transitioning to controlled experiments within our custom network.

\textbf{I. Behavior Baselining.} The first phase of our connection methodology involves observing how the Tesla vehicle interacts with commercial cellular networks, allowing us to establish a baseline reference for expected network behaviors. This includes analyzing PLMN selection, roaming policies, network attachment sequences, and cell reselection priorities. Even though black-boxed, by monitoring its automatic network selection process, we can still infer some key connectivity patterns and restrictions. For example, we discovered that the vehicle contains a T-Mobile eSIM profile, thus it prioritizes T-Mobile access in a given area. Additionally, we identified (1) \textit{Continuous Cell Selection}: The TCU scanned and reattached immediately when released from isolation (e.g., Faraday cage), suggesting no long attachment backoff under normal conditions. (2) \textit{eSIM Dominance}: The pre-provisioned eSIM takes priority even in the presence of an active physical SIM. (3) \textit{Dual Connectivity}: LTE remains active alongside WiFi, likely for telemetry. (4) \textit{PLMN Filtering}: The TCU only attempted to roam onto operator-allowed PLMNs, not all visible ones. (5) \textit{Retry Logic}: After repeated failed attachments, a cooldown was enforced, temporarily suspending retries.

\textbf{II. Use of Cell Mapping.} At this phase, we performed cell mapping and passive network reconnaissance in the area to identify active networks, extract key configuration parameters (e.g., Tracking Area Codes (TACs), EARFCNs, signal metrics (RSRP, RSRQ, SINR), serving PLMNs, LTE bands), and determine a suitable strategy for our custom LTE setup. This step was essential for ensuring compatibility with the Tesla TCU while avoiding interference with commercial networks. To survey the local cellular environment, we utilized cell mapping applications and radio scanning tools that allowed us to log critical network parameters, specifically (1) \textit{CellMapper}~\cite{CellMapper}, (2) \textit{srsRAN}~\cite{srsRAN} and (3) \textit{Android Dialer Service}~\cite{AndroidDialer} with the Android engineering menu to extract real-time information.

\textbf{III. SIM Card Programming.} We used multiple SIM cards to support different phases of experimentation. Commercial SIMs (e.g., T-Mobile, AT\&T) were used for baseline behavior analysis for the vehicle and smartphones. An Amarisoft test SIM (PLMN 00101) enabled attachment testing, while a Sysmocom programmable SIM was configured to impersonate a real network (e.g., T-Mobile, PLMN 310260) and used for full authentication with our LTE core.

For network authentication, we programmed our own set of security credentials, including Authentication Key (Ki), Operator Code (OPC), Access Point Names (APN) and PLMN settings. The same authentication parameters were also reflected on the network side, ensuring that our LTE core recognized the SIM as a legitimate subscriber within our controlled environment. Consequently, to configure the SIM cards, we utilized pySim~\cite{pySim} and a smart card reader. To read we used \texttt{\$./pySim-read.py -p0}, while to write \texttt{\$./pySim-prog.py -p 0 -a <Admin Code> -x <MCC> -y <MNC> -i <IMSI> -s <ICCID> -o <OPC> -k <Key>}. Once the SIM was programmed, it was inserted into the Tesla’s SIM slot, and the eSIM was disabled via Tesla Toolbox. We did not encounter any authentication or compatibility issues.

\textbf{IV. Custom LTE Network Preparation.} In this phase, our primary objective was to validate and fine-tune the custom LTE network setup to ensure correct attachment and operation before attempting to connect the Tesla vehicle. To achieve this, we leveraged the data from previous phases and employed standard LTE-enabled smartphones (iPhone and Android) as reference devices. By using them, we performed manual and automatic network selection tests to ensure that our network configurations were applied correctly. With the correct APN settings and the airplane mode toggled to trigger reattachment, the phones successfully connected every time and were able to access the internet through our core network. Even without a SIM card, the devices detected and attempted to connect to our LTE base station.

Next, to ensure a stable and reliable network environment, we performed signal measurements using standard LTE key performance indicators (KPIs), including the RSSI, RSRP, and SINR. We did not encounter any network stability issues, and the devices were able to access multimedia content online, indicating that the network provided sufficient resources for the vehicle testing subsequently.

Finally, we confirmed full EPC and eNodeB functionality through: (i) persistent registration and session retention, (ii) web access confirming end-to-end PDU session establishment, (iii) correct PLMN detection in manual and automatic scan modes, and (iv) stable reconnection after disconnections and successful handovers/reattachment. These baseline checks ensured readiness for controlled integration with the TCU.

To confirm that our EPC and eNodeB setup was fully operational, we validated: 

\begin{enumerate}
    \item \textit{Successful network attachment of reference devices}. Phones could register, authenticate, and most importantly remain connected to the network (no release or connection drop).
    \item \textit{Internet accessibility}. Connected phones were able to browse the web, confirming correct APN, IP configurations, and PDU session establishment.
   \item \textit{PLMN visibility and selection}. The custom PLMN was detected also in manual scan mode (apart from the automated attachment), ensuring it was correctly broadcasted.
   \item \textit{Handover and reattachment stability}. Devices were toggled in and out of airplane mode to test whether they could reliably reconnect without failure. Reattachments were tested to ensure that devices could reconnect back after potential failures.
\end{enumerate}

Once these validations were complete, we confirmed that the LTE network setup was functioning properly and that the Tesla should, based on all tested parameters, be able to connect without issues. With this empirical verification, we proceeded to the final stage—connecting and integrating the Tesla vehicle with our custom LTE network.

\textbf{V. Configuring the Vehicle.} Multiple steps were required to physically install the SIM, access the vehicle’s diagnostic interfaces, and apply necessary system reboots to enforce network selection changes. As Tesla vehicles do not provide a user-facing SIM slot (depending on the model), it was necessary to physically access the TCU to insert our programmable SIM card, involving the following steps:

\begin{enumerate}
\item Removed the headliner to access the SIM slot.

\item Shut down the Tesla and disconnect the 12V battery. Since the TCU remains powered even after the vehicle is off, the 12V battery under the rear seat was disconnected to ensure a complete power cycle before inserting the new SIM.

\item Inserted the SIM card securely into the slot.

\item Reconnected the 12V battery and waited for the power relay to ``click". This step ensured that all vehicle subsystems, including the TCU, received a proper reset after the SIM swap.

\item Turned on the Tesla and allowed the system to initialize. The vehicle’s network stack was then given time to attempt attachment.
\end{enumerate}

``Hard Reboots" (Full Power Cycle), which mean disconnecting and reconnecting the 12V battery, were the only reliable method to enforce a complete network reset. However, hard reboots were not frequent in our experimentation, only when the SIM cards were inserted. We mostly leveraged ``Soft Reboots", meaning turning off/on the vehicle, even though we did not witness any significant effect on the network connection. This indicates that it can still be connected to a network, although it may seem turned off.

\begin{figure}[!t]
     \centering
     \includegraphics[width=\columnwidth]{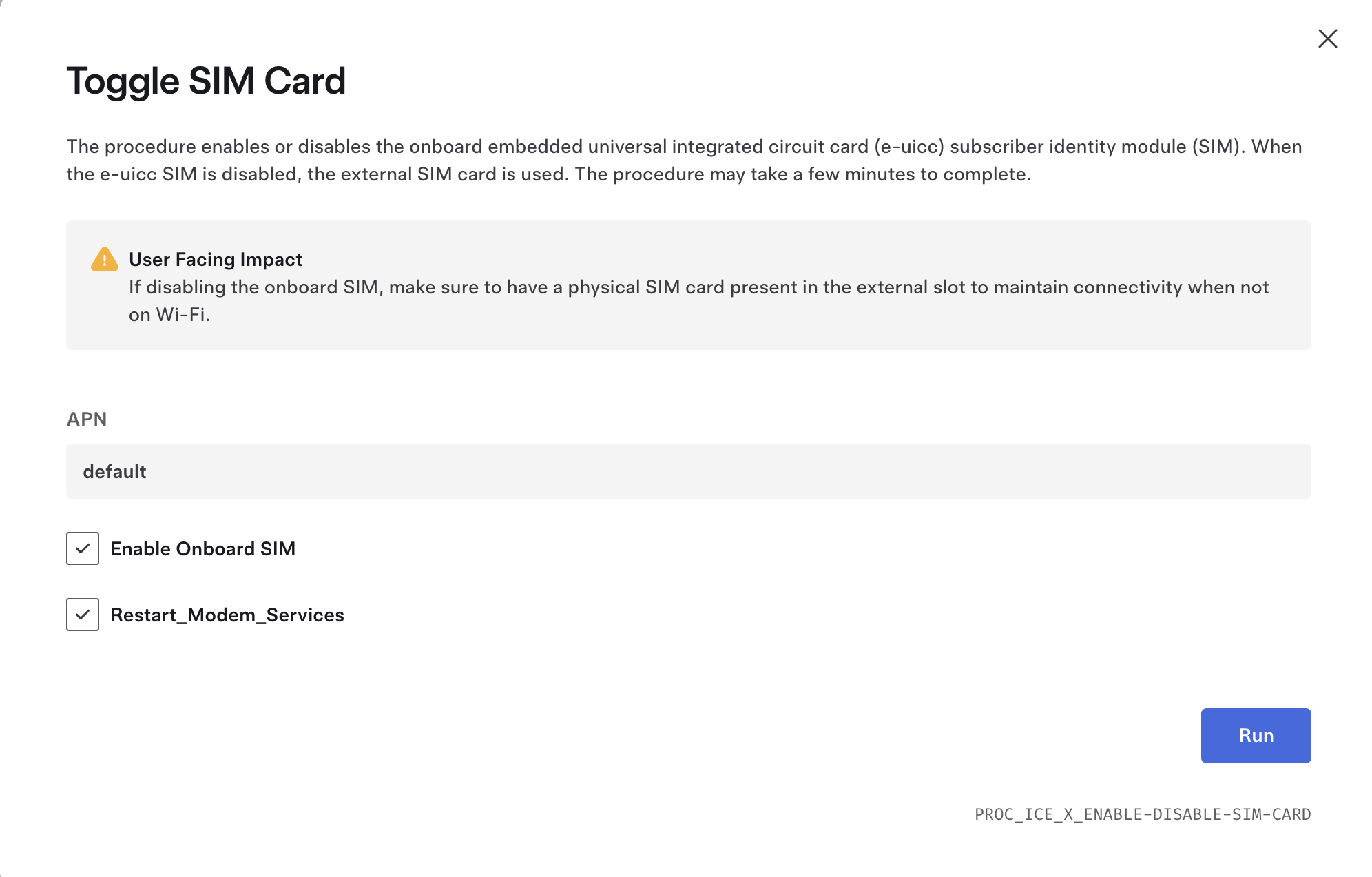}
     \caption{SIM Activation in the Tesla Toolbox.}
     \label{fig:toolbox-toggle}
\end{figure}

\begin{figure}[!t]
     \centering
     \includegraphics[width=\columnwidth]{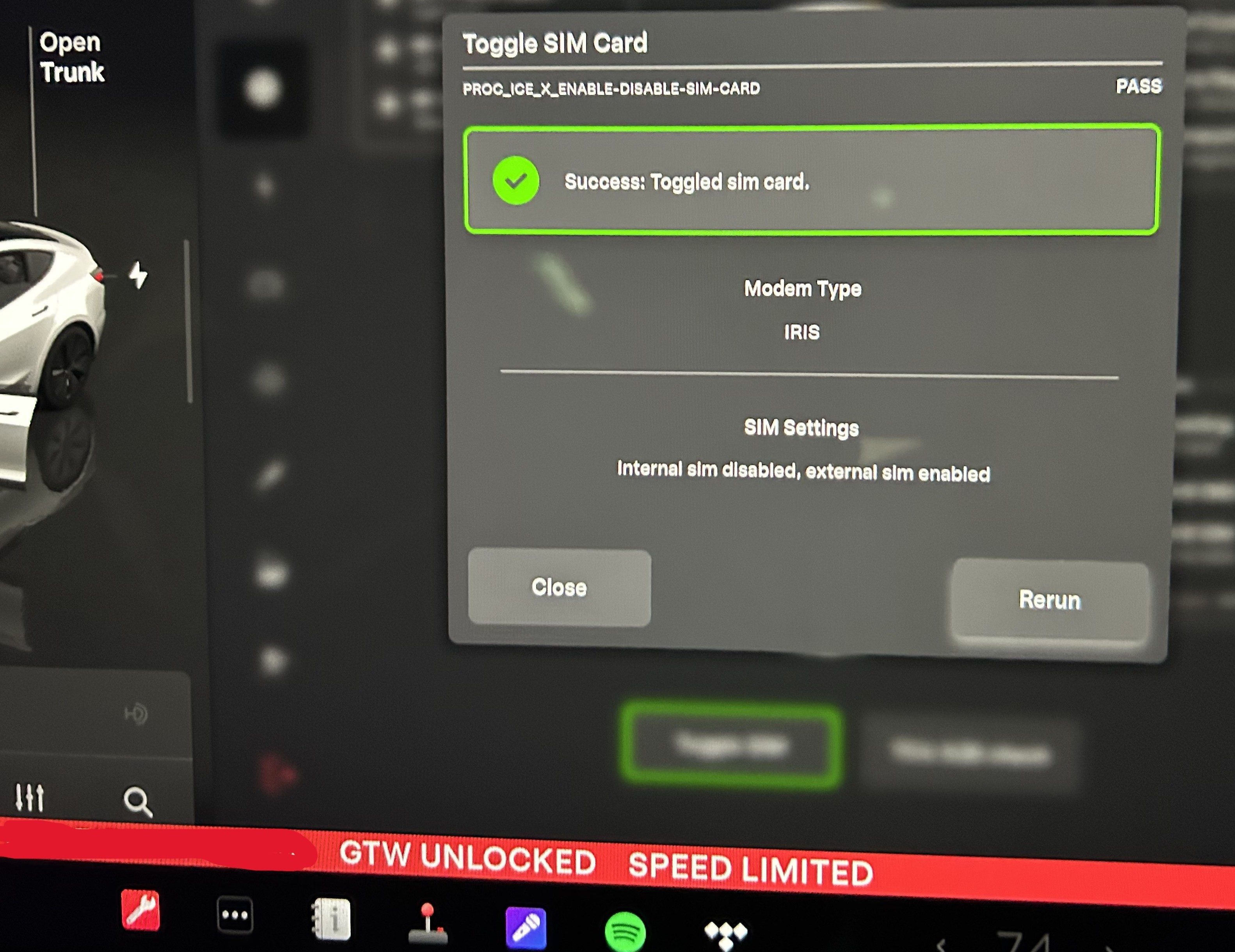}
     \caption{Successful physical SIM card activation.}
     \label{fig:lte-sim}
\end{figure}

\textit{Accessing Configurations \& Debugging.} We utilized the Tesla Toolbox to monitor and control the network selection process. By default, Tesla prioritizes its internal eSIM over the physical SIM slot. Using Tesla Toolbox, we accessed the network configuration settings and explicitly disabled eSIM functionality, forcing the vehicle to utilize the physical SIM. Figure~\ref{fig:toolbox-toggle} shows this action in Toolbox, even though it can also be performed on the Control Screen with ``Service Mode Plus''. Additionally, Figure~\ref{fig:lte-sim} illustrates the successful activation of a SIM card on the Control Screen.

\begin{figure}[!ht]
     \centering
     \includegraphics[width=\columnwidth]{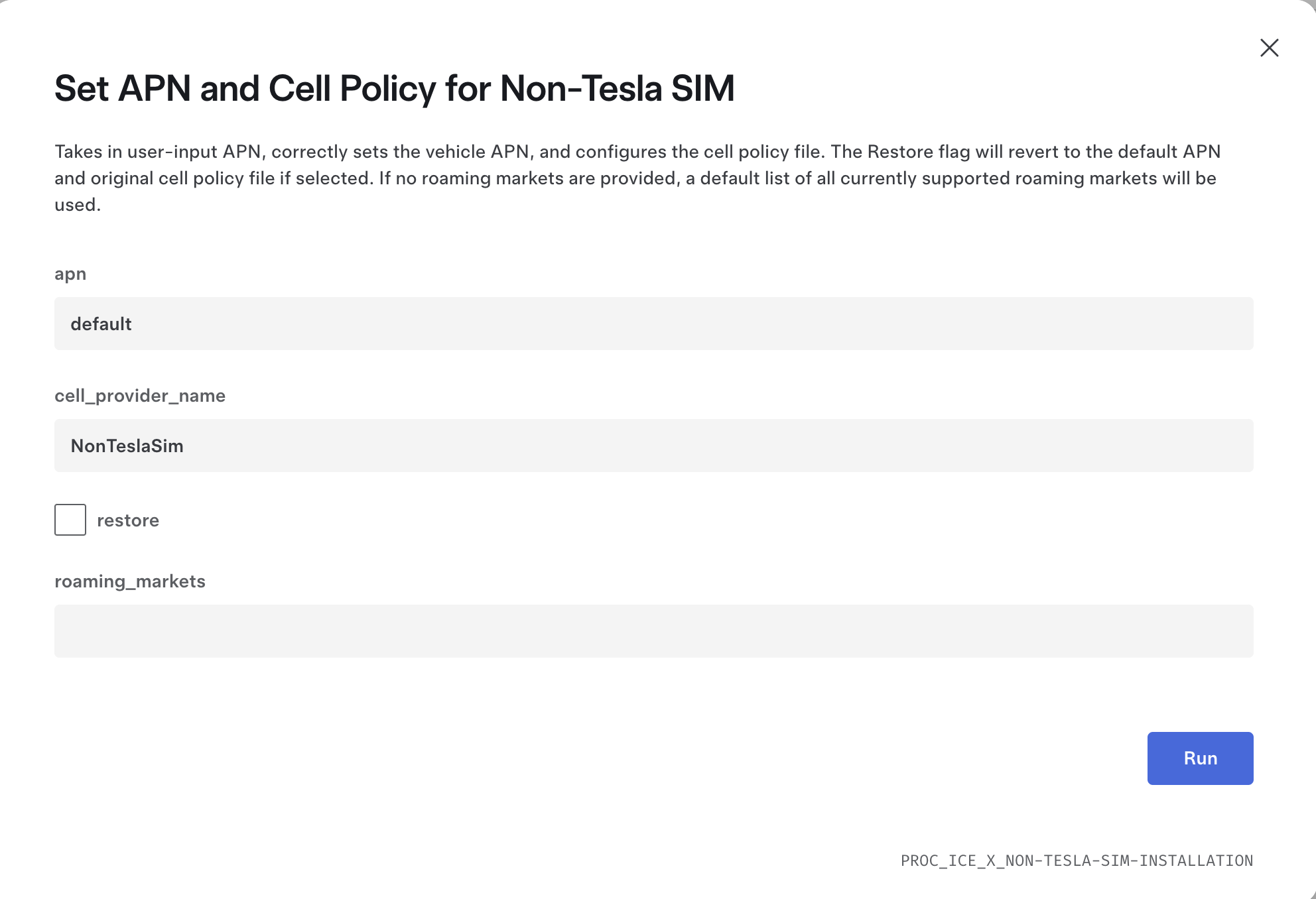}
     \caption{APN parametrization in the Tesla Toolbox.}
     \label{fig:toolbox-apn}
\end{figure}

\begin{figure}[!b]
     \centering
     \includegraphics[width=\columnwidth]{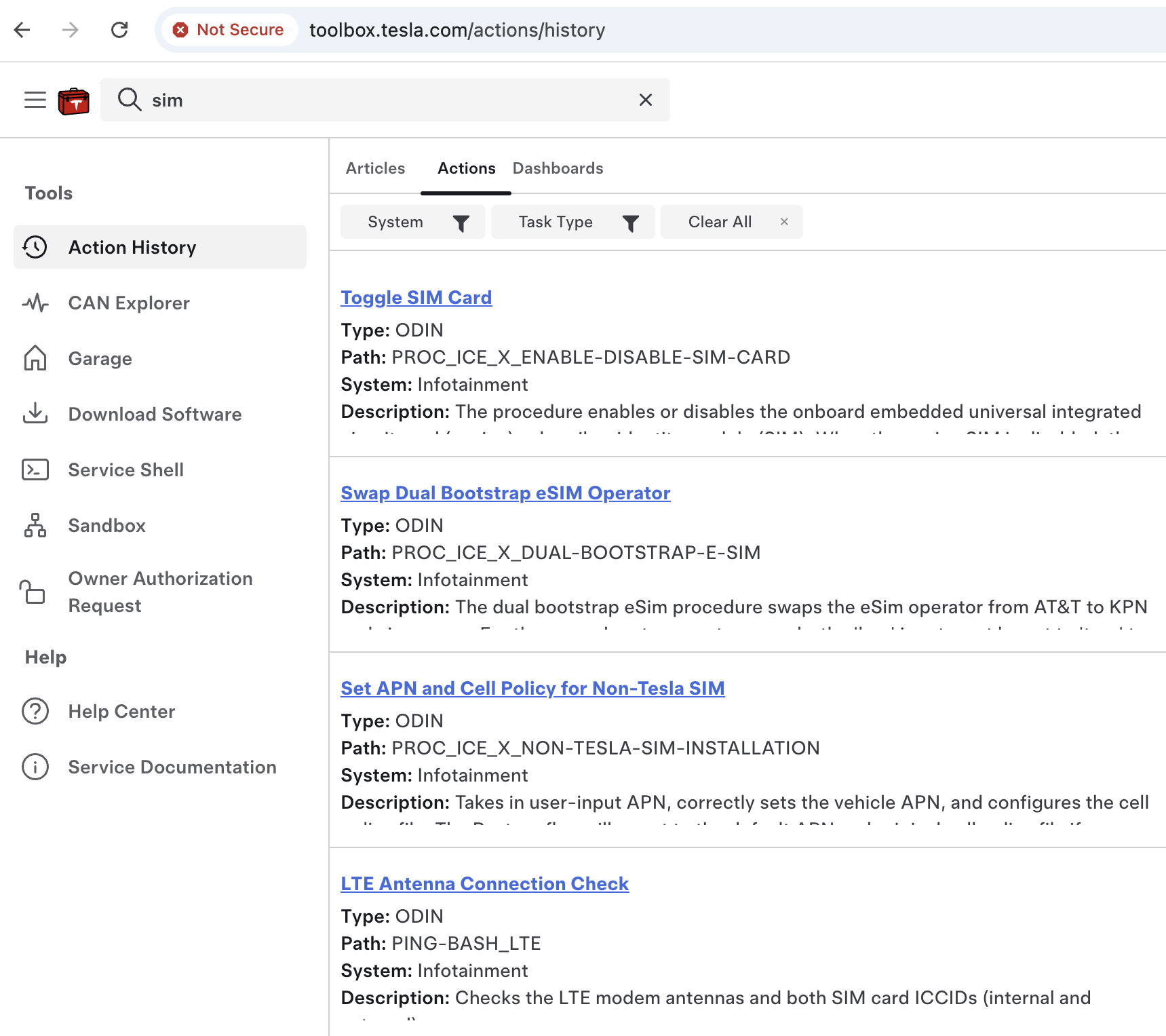}
     \caption{Example of Tesla Toolbox Features for Cellular Connectivity.}
     \label{fig:toolbox-features}
\end{figure}

\begin{figure}[!ht]
     \centering
     \includegraphics[width=\columnwidth]{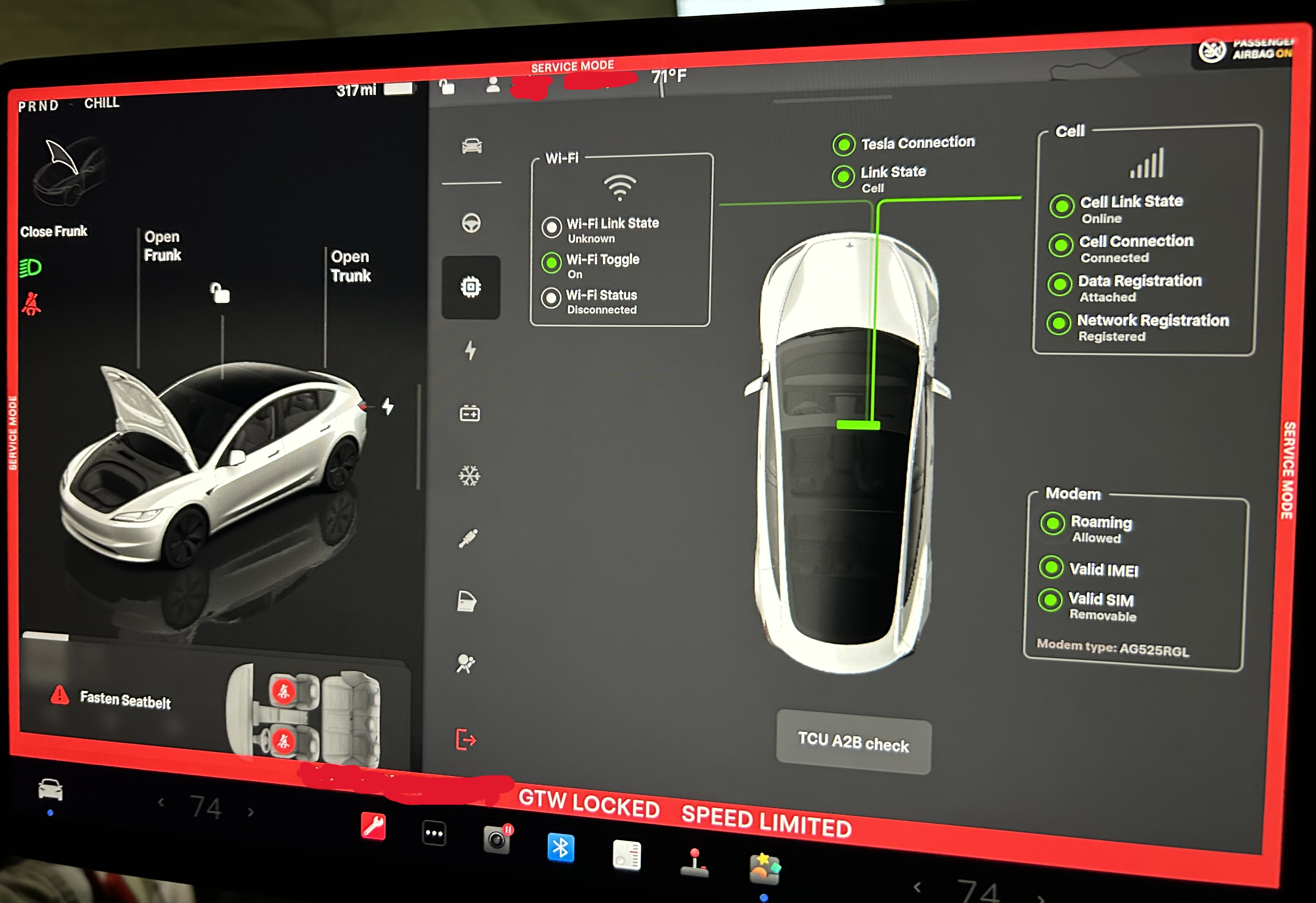}
     \caption{Successful Model 3 LTE connection.}
     \label{fig:lte-wifi}
\end{figure}

\begin{figure}[!ht]
     \centering
     \includegraphics[width=0.8\columnwidth]{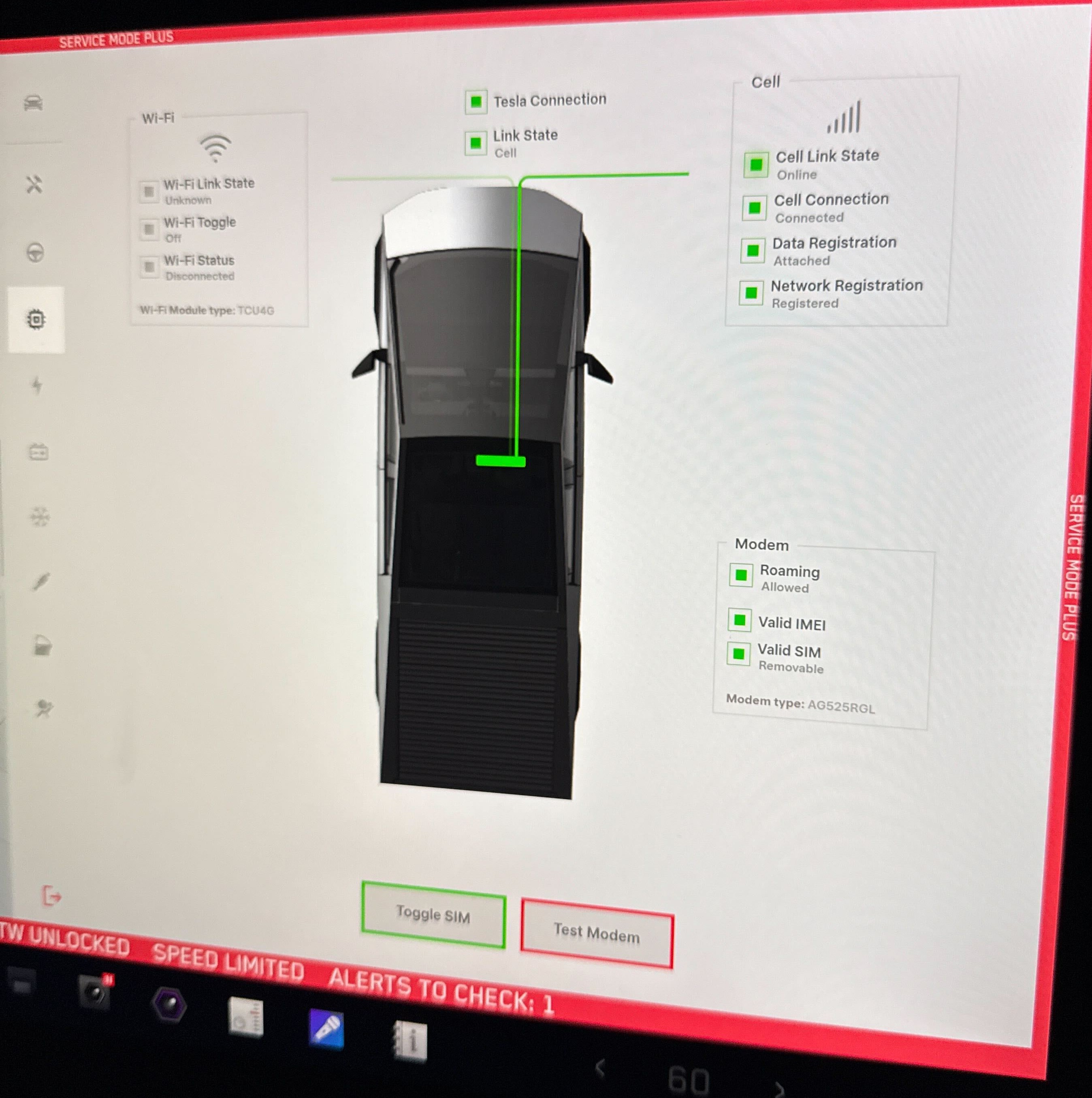}
     \caption{Successful Cybertruck LTE connection.}
     \label{fig:lte-cybertruck}
\end{figure}

The Toolbox interface allowed us to confirm that the TCU detected our custom-programmed SIM and that the registration requests were being sent to the network. We cross-referenced these logs with our LTE core network logs to ensure that the Tesla was successfully initiating attachment requests. It should be mentioned that the Tesla vehicles include the ``Service Mode", accessible through the infotainment system, from which we can verify some connection steps of the ``modem" (i.e., Valid SIM, Valid IMEI, Roaming) and ``cell" (i.e., Network Registration, Data Registration, Cell Connection, Cell Link State), as shown in Figures~\ref{fig:lte-wifi} and~\ref{fig:lte-cybertruck}. While the Tesla Toolbox and Service Mode provided useful insights, they did not allow manual network selection specifically in the same way as in smartphone devices. We used Toolbox further for correct network parametrization, such as the APN. As a result, network-side configurations had to be carefully aligned to match Tesla’s expected settings.

Figures~\ref{fig:lte-wifi} and~\ref{fig:lte-cybertruck} depict the successful connection of the Tesla vehicle to our custom LTE network with Amarisoft. Using netstat in the interactive shell, we verified that there are open connections related to the modem as well. With the successful attachment confirmed, we monitored Tesla’s network traffic to ensure its stability and reliability.

\section{Artifact Available}

An artifact associated with this paper is publicly available on Zenodo at~\url{https://doi.org/10.5281/zenodo.19699792}. It includes packet captures collected during our experiments, covering representative LTE control-plane signaling and data-plane communication between the evaluated vehicles and backend services. These traces provide independently inspectable evidence for the observations reported in Section~5, including the advertised cellular capabilities and the disruption of backend connectivity during the false base station experiments. Full replication of the experiments requires specialized infrastructure, licensed cellular-network software, access to the evaluated vehicles, and an RF-isolated testing environment.

\begin{figure}[!ht]
     \centering
     \includegraphics[width=\columnwidth]{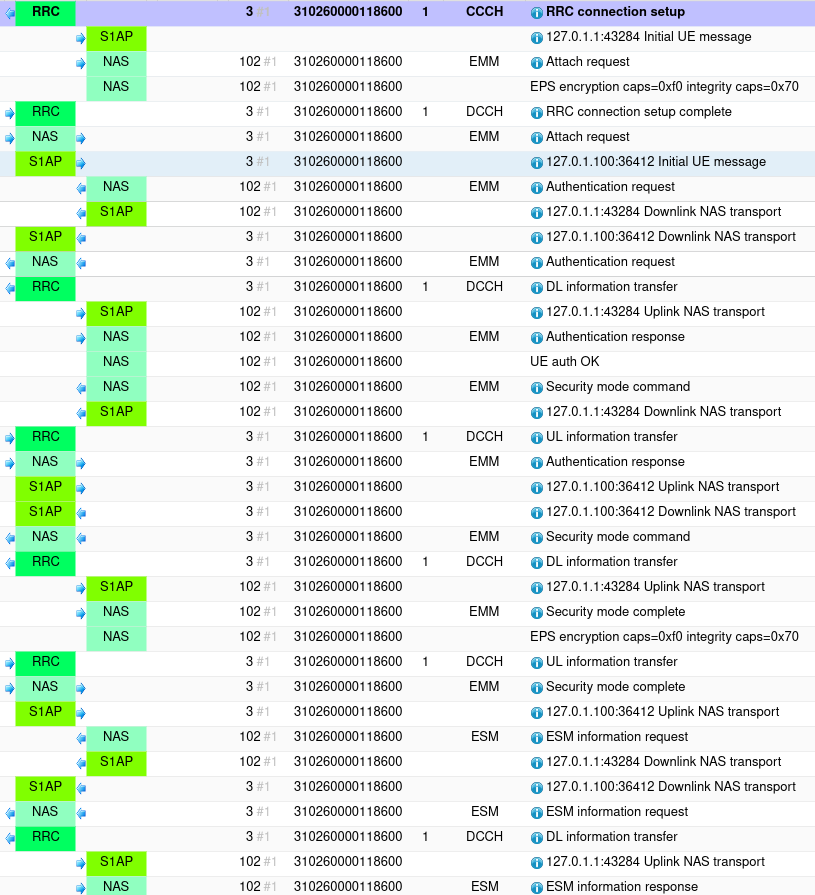}
     \caption{Normal LTE connection to our Amarisoft.}
     \label{fig:lte-normal-connection}
\end{figure}

\begin{figure}[!ht]
     \centering
     \includegraphics[width=\columnwidth]{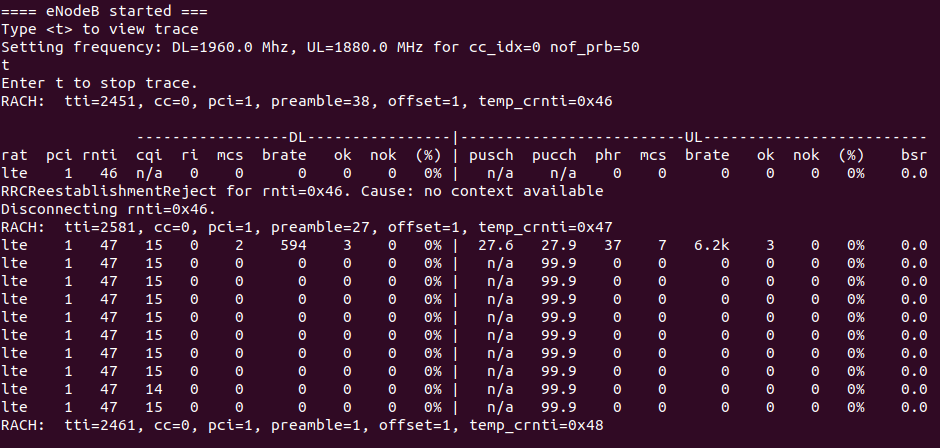}
     \caption{RACH and RRC connection attempts from the attacker's perspective.
     }
     \label{fig:lte-fbs-attacker}
\end{figure}

\begin{figure}[!ht]
     \centering
     \includegraphics[width=\columnwidth]{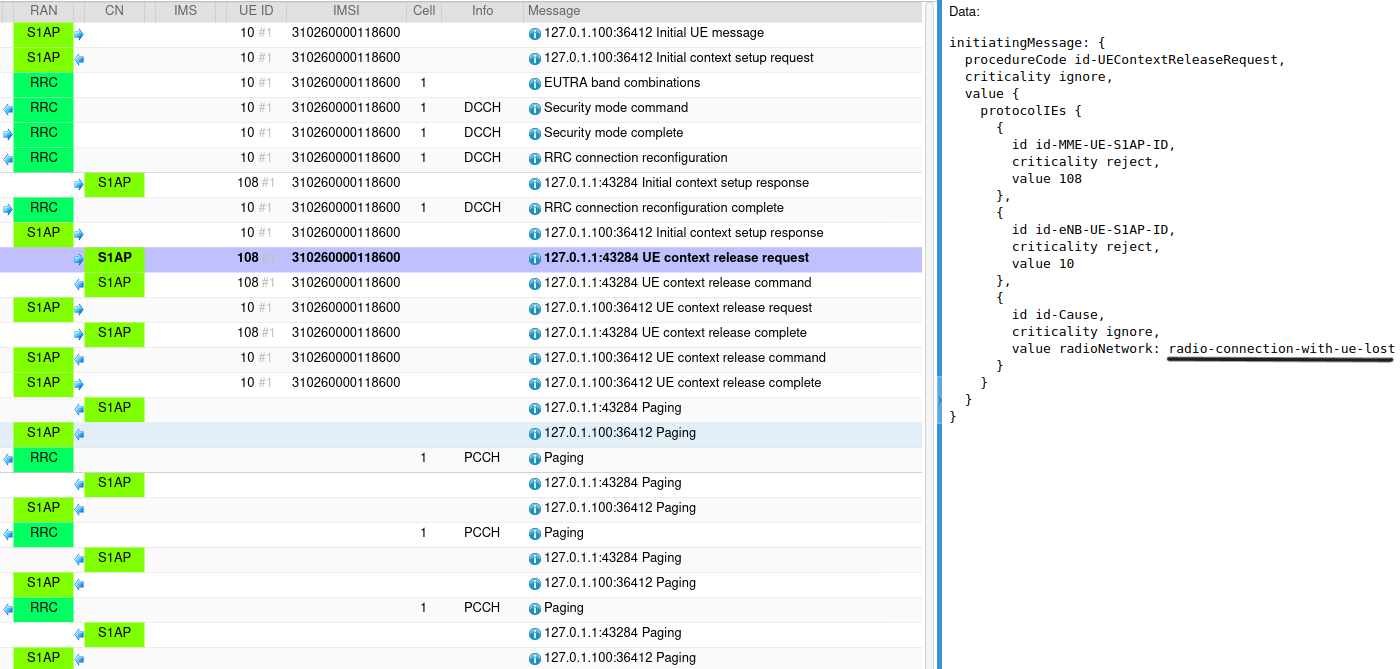}
     \caption{Network's perspective during the FBS attack. The UE disappears and cannot be reached by the network.}
     \label{fig:lte-fbs-net}
\end{figure}

\end{document}